\documentclass{aa}
\usepackage{graphicx}
\usepackage{txfonts}
\usepackage{subcaption} 
\pdfoutput=1 
\usepackage{amssymb}
\usepackage{gensymb}
\usepackage[caption=false]{subfig}
\usepackage{booktabs}
\usepackage{xcolor}
\usepackage{multirow}
\usepackage{xspace}
\usepackage{amsmath}
\usepackage{lscape}             
\usepackage{placeins}  

\newcommand{\Teff}{$T_{\mathrm{eff}}$}
\newcommand{\modelone}{$\mathcal{M}1$}
\newcommand{\modeltwo}{$\mathcal{M}2$}
\newcommand{\modelthree}{$\mathcal{M}3$}
\newcommand{\taugran}{$\tau_{\mathrm{gran}}$}

\newcommand{\sigmagran}{$\sigma_{\mathrm{gran}}$}

\newcommand{\numax}{$\nu_{\mathrm{max}}$}
\newcommand{\sigmaenv}{$\sigma_{\mathrm{env}}$}

\begin{document}

    \title{Granulation in Color: The Chromatic Stellar Flicker of LTT-9779 and WASP-121}

    \author{William~C.~Waalkes\inst{1}\corrauth{waalkes@mps.mpg.de}
    \and Nadiia Kostogryz\inst{1}\email{kostogryz@mps.mpg.de}
    \and Valeriy Vasilyev\inst{1}\email{vasilyev@mps.mpg.de}
    \and Rafael Luque\inst{2}\email{rluque@iaa.es}
    \and Alexander Shapiro\inst{3,1}\email{shapiroa@mps.mpg.de}
    \and Jordi Blanco-Pozo\inst{4,5,6}\email{blanco@ice.csic.es}
    \and Ana Glidden\inst{7,8}\email{ana.glidden@cfa.harvard.edu}
    \and Jeff Valenti\inst{9}\email{valenti@stsci.edu}
    \and Sami Solanki\inst{1,10}\email{solanki@mps.mpg.de}
    \and Ignasi Ribas\inst{4,5}\email{iribas@ice.csic.es}
    \and Sara Seager\inst{11}\email{profseager@mit.edu}
    \and Laurent Gizon\inst{1}\email{gizon@mps.mpg.de}
    }

    \institute{Max-Planck-Institut f\"ur Sonnensystemforschung, Justus-von-Liebig-Weg 3, 37077 G\"ottingen, Germany
    \and Instituto de Astrof\'isica de Andaluc\'ia (IAA-CSIC), Glorieta de la Astronom\'ia s/n, 18008 Granada, Spain
    \and University of Graz
    \and Institut de Ci\`encies de l'Espai (ICE, CSIC), Campus UAB, C/Can Magrans s/n, 08193 Bellaterra, Spain
    \and Institut d'Estudis Espacials de Catalunya (IEEC), C/Esteve Terradas, 1, Edifici RDIT, Campus PMT-UPC, E-08860 Castelldefels (Barcelona), Spain
    \and Facultat de F\'isica, Universitat de Barcelona (UB), Mart\'i Franqu\`es 1, 08028, Barcelona, Spain
    \and Center for Astrophysics, Harvard \& Smithsonian, 60 Garden Street, Cambridge, MA 02138, USA
    \and NSF AAPF
    \and Space Telescope Science Institute, 3700 San Martin Dr., Baltimore, MD 21218, USA
    \and School of Space Research, Kyung Hee University, Yongin, Gyeonggi 17104, Republic of Korea
    \and Canadian Institute for Theoretical Astrophysics, University of Toronto, 60 ST.George Street, Toronto, ONM5S3H8, Canada
    }
    
    \abstract
    {Granulation is a persistent feature of stars with outer convective layers and is the photospheric signature of near-surface convective energy transport.
    In disk-integrated observations, it produces stochastic brightness variations that contribute substantially to stellar variability on timescales of minutes to hours.
    High-precision, spectrally resolved time-series observations with JWST now provide a new opportunity to study the wavelength-dependent signatures of stellar granulation.}
    {This work aims to measure the amplitude and characteristic timescale of granulation-induced variability for solar-type stars and to investigate whether the granulation signal shows significant chromatic variation in the near-infrared.}
    {We accessed publicly available JWST/NIRISS SOSS time-series observations of LTT-9779 (G7V) and WASP-121 (F6V) and de-trended the light curves by fitting the transits, eclipses, and planetary phase curves. 
    We computed power spectra from the de-trended light curves and fit them with a Harvey granulation profile.
    We first analyzed the broadband light curves to measure and fix the granulation timescale and then repeated the power-spectrum analysis in individual wavelength channels to measure the chromatic dependence of the granulation amplitude.}
    {For LTT-9779, we measure a broadband granulation amplitude of \sigmagran$=53.8^{+4.1}_{-4.0}$ppm at a timescale of \taugran$=4.8^{+1.5}_{-1.3}$min. For WASP-121, we measure \sigmagran$=59.3\pm4.3$ppm and \taugran$=6.3^{+2.1}_{-1.8}$min.
    We also measure peak oscillation frequencies of \numax$=3382^{+166}_{-191}\mu$Hz and $1649^{+55}_{-69}\mu$Hz in the bandpass-integrated power spectra of LTT-9779 and WASP-121, respectively.
    Furthermore, we recover chromatic granulation amplitudes that show an overall decrease in amplitude towards longer wavelengths.}
    {We present an analysis of chromatic granulation signatures observed by JWST.
    Our results demonstrate that JWST time-series observations can be used to characterize stellar granulation and to measure its chromatic dependence.
    This opens a new observational window on short-timescale stellar variability and provides a way to characterize chromatic stellar noise in transit light curves, improving the robustness of exoplanet measurements and offering a complementary diagnostic of stellar photospheric properties. }

    
    \maketitle

\nolinenumbers

\section{Introduction} \label{intro}

Granulation is the surface phenomenon associated with near-surface convective energy transport in stellar photospheres \citep[e.g.,][]{Nordlund2009}.
On the Sun, this phenomenon was first reported through visual observations by \citet{Herschel1801} and others, and later captured in detail through drawings \citep{Secchi1875} and short-exposure photography \citep{Janssen1877,Janssen1896}. Solar granulation subsequently became a key observable signature of photospheric convection.
Granulation produces stochastic brightness variations as hot plasma rises, cools radiatively near the stellar surface, and descends through darker intergranular lanes \citep{Dravins1982,Nordlund2009}.
Stars also support acoustic oscillations excited by granulation, contributing a stochastic variability on the timescale of minutes \citep{Leighton1962}. 
In disk-integrated observations, these phenomena introduce stellar variability with characteristic amplitudes and timescales.
For low-activity stars like the Sun, granulation and oscillations are among the dominant contributors to disk-integrated photometric variability on timescales of minutes to hours, while magnetic activity becomes increasingly important on rotational and longer timescales \citep{Bastien2013,Shapiro2017}.
These convective processes can hinder the detection and characterization of exoplanets \citep{Aigrain2004,Cegla2019,Meunier2020} by raising the stellar noise floor and creating a stochastic stellar ``flicker''.

Flicker affects exoplanet characterization through both spectroscopic and photometric channels.
In high-resolution spectroscopy, it alters spectral-line shapes, asymmetries, and wavelength shifts, introducing time- and wavelength-correlated radial-velocity (RV) variations that can obscure the mass measurement of Earth-like planets orbiting Sun-like stars \citep[e.g.,][]{SchrijverZwaan2000,Guo2022,Frame2025,Burt2025}.
Simulations of RV surveys indicate that granulation can increase the uncertainty of Earth-analogue mass measurements by approximately a factor of three compared with a case without stellar variability \citep{Luhn2023}. In transit observations, neglecting time-correlated stellar flicker can introduce errors of up to 10$\%$ in the inferred planet-to-star radius ratio for an Earth-sized planet orbiting a Sun-like star \citep{Sulis2020}.
The wavelength dependence of this variability can also affect transit depths measured in individual spectral channels and, consequently, the inferred planetary transmission spectrum.
Recent transit studies with JWST's NIRISS/SOSS instrument \citep{Doyon2023} have found signals attributed to granulation with characteristic amplitude of $\sim$50-100 ppm in the light curves of main sequence stars \citep{Cadieux2024,Coulombe2025,Splinter2025,Pass2026}. 
With the extreme precision of space telescopes and ground-based RV spectrographs, accounting for stellar variability has become an essential aspect of exoplanet discovery and characterization.

Additionally, granulation and acoustic oscillations are important observable phenomena for constraining fundamental stellar parameters and understanding the physics of stellar interiors \citep[][]{Ness2018,Aerts2021}.
Granulation amplitudes and timescales depend on the fundamental stellar parameters; effective temperature ($T_{\mathrm{eff}}$), surface gravity (log$g$), and metallicity ($[\rm {M/H}]$), as well as magnetic activity \citep{Trampedach1998,Bahcall2006,Ludwig2009,Fabbian2009,Mathur2011,KjeldsenBedding2011,Kallinger2014A&A...570A..41K,Cranmer2014,Bastien2016,Corsaro2017,Tayar2019,RodriguezDiaz2022}.
The intensity of the granulation pattern scales with $T_{\mathrm{eff}}$ and $[\rm {M/H}]$ while the number of granules on the surface of the star, N$_{\mathrm{granules}}$, scales with log$g$.
The disk-integrated amplitude of stellar flicker scales directly with the intensity contrast of the granulation pattern and inversely with the square root of the number of independently varying granules $1/\sqrt{N_{\mathrm{granules}}}$.
In other words, the observed effect decreases for cooler stars because they have a stronger surface gravity and more granules.

\citet{Cranmer2014} suggested that increased surface magnetism would reduce the magnitude of stellar flicker because magnetic fields suppress convection.
Using three-dimensional radiative-MHD simulations, \citet{Kostogryz2026} studied the wavelength-dependence of granulation for K2, G2, and F3 stellar models, including G2 models with different metallicities.
They showed that magnetic fields decrease the intensity contrast of granulation by suppressing convective temperature fluctuations while also enhancing the intensity contrast from localized magnetic structures, especially near the limb.
In disk-integrated spectra, these effects compete with each other to modulate the observed variability depending on stellar type, metallicity, and wavelength.
Thus, chromatic granulation provides a promising diagnostic of stellar photospheric properties, surface magnetism, and near-surface convection.

The vast majority of stellar granulation research has focused on single-bandpass light curves of evolved stars, with only recent work by \citet{Sreenivas2025} taking advantage of TESS and Kepler light curves to perform a two-channel chromatic analysis of granulation and oscillation and demonstrating evidence of the predicted decrease in granulation amplitude with wavelength \citep{KjeldsenBedding1995,Lund2019}.
Studying the wavelength-dependence of convective processes can help us clarify their driving mechanisms and provide a systematic improvement to the study of exoplanet transmission spectra. 
Here, we aim to detect and characterize spectral granulation signatures in continuous high-cadence time-series spectroscopy from JWST NIRISS/SOSS.
This work starts a series on ``Granulation in Color" (AR \#11527 (PIs: J. Valenti \& A. Glidden) which will use the unprecedented spectral sensitivity and photometric precision of JWST to compare wavelength-dependent stellar flicker with 3D radiative MHD models.

This paper is laid out as follows: in Section \ref{section:methods}, we describe the target selection process and how we reduced and conditioned the JWST observations in order to study their power spectra.
In Section \ref{section:results}, we present the results of Bayesian inference on the power spectra, which lead to measurements of the granulation timescale and amplitude. We further compare the chromatic variation in amplitude to theoretical predictions from stellar granulation simulations. 
Finally, in Section \ref{discussion}, we place these results in the context of previous studies of solar and stellar granulation. 

\begin{figure*}
    \centering
    \includegraphics[width=0.45\linewidth]{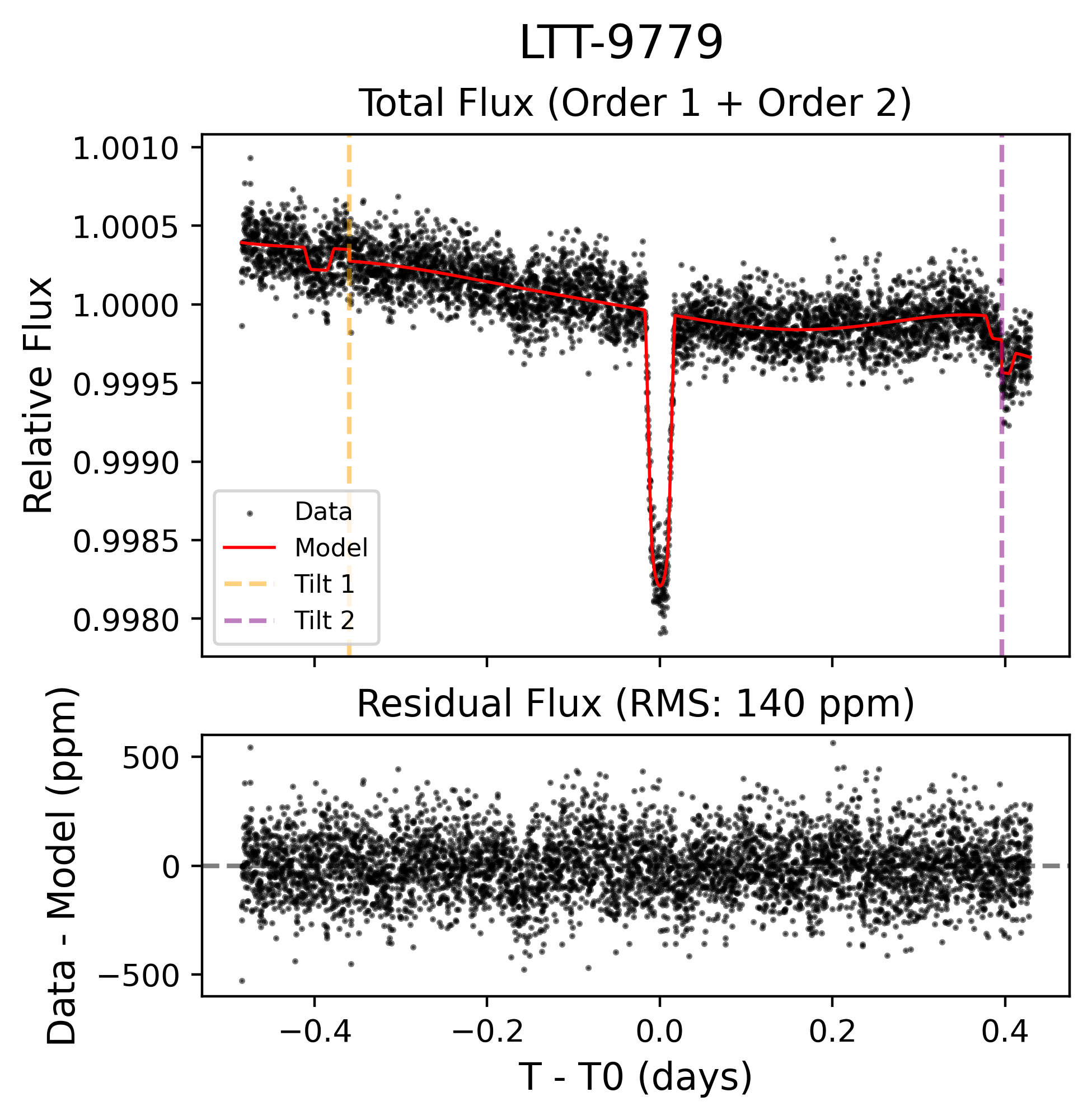}
    \includegraphics[width=0.45\linewidth]{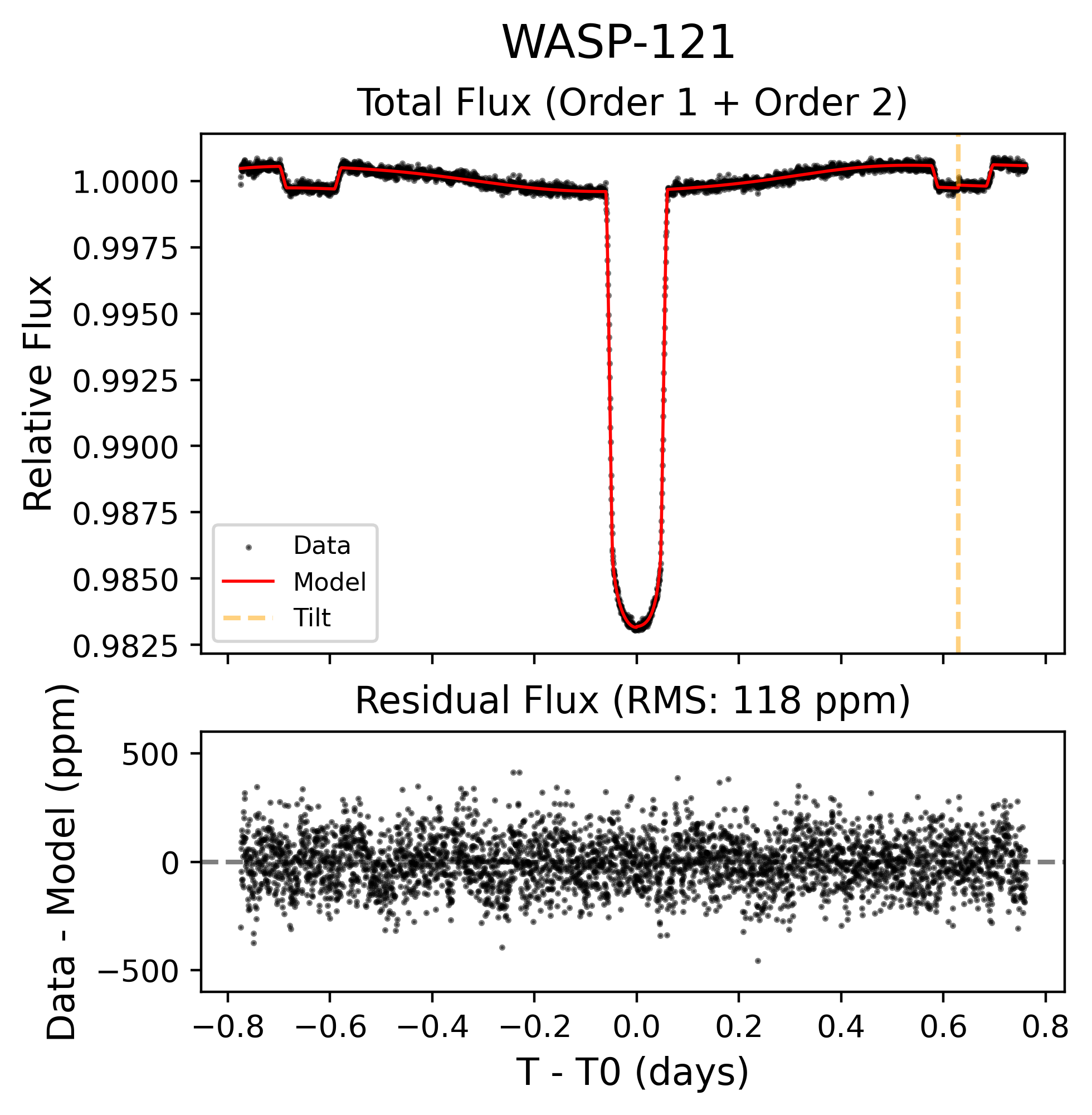}
    \caption{Bandpass-integrated (BI) light curves with transit and eclipse models (top row) for LTT-9779 (left) and WASP-121 (right). Also shown (bottom) is the associated de-trended light curve created by subtracting the model (red line) from the data; this residual time series contains information about the stellar granulation and oscillations. Tilt events are instrumental offsets related to the movement of JWST's mirrors.
    }
    \label{fig:transit-models}
\end{figure*}

\begin{table}[h!]
    \centering
    \begin{tabular}{lcc}
         Parameter                  &   LTT-9779        &   WASP-121        \\
         \hline
         \Teff~[K]                  &   $5445\pm84$     &   $6628\pm66$     \\
         $\rm M_*$~[M$_{\odot}$]    &   $1.03\pm0.04$   &   $1.385\pm0.016$   \\
         $\rm R_*$~[R$_{\odot}$]    &   $0.95\pm0.01$   &   $1.461\pm0.005$ \\
         $\rm M/H$~[dex]            &   $0.25\pm0.08$   &   $0.17\pm0.05$   \\
         log$g$~[cgs]               &   $4.43\pm0.31$   &   $4.251\pm0.003$\\
         Cadence [s]                &   16.5            &  38.5 \\
         Baseline~[hr]              &   20.6            &  36.9 \\ 
         \hline
    \end{tabular}
    \caption{Details for the stellar parameters and timeseries baseline and cadence for the two targets in this study. LTT-9779 parameters are taken from \citet{Jenkins2020} and WASP-121 parameters are from \citet{Sing2024}.}
    \label{tab:target-parameters}
\end{table}

\section{Methods} \label{section:methods}

\subsection{Target Selection and Data Reduction}
\label{subsection:methods-data}

\subsubsection{LTT-9779 and WASP-121}
We chose these targets from the list of publicly available JWST transit observations compiled in the Transiting Exoplanets List of Space Telescope Spectroscopy (TrExoLiSTS) database \citep{Nikolov2023}.
Starting with the full list of all publicly available transmission spectrum observations from JWST, we narrowed the search to F- and G-type stars as their flicker amplitudes are greater than those of cooler stars.
From those observations we selected our final targets as those with the longest baselines -- NIRISS/SOSS phase curve observations of LTT-9779 (21-hour baseline with 16.5s cadence) and WASP-121 (37-hour baseline with 38.5s cadence). 
Instrumental and stellar parameters relevant to this study are listed in Table \ref{tab:target-parameters}.

The NIRISS/SOSS observations of WASP-121 and LTT-9779 analyzed in this work are both phase curves obtained as part of the JWST GTO Program 1201 \citep{Lafreniere2017}.
The observations of WASP-121 were taken in October 2023 using the SUBSTRIP256 subarray and published in \citet{Splinter2025}.
The observations of LTT-9779 were taken in July 2022 using the SUBSTRIP256 subarray and published in \citet{Coulombe2025}.
We downloaded the observations from the Mikulski Archive for Space Telescopes (MAST) and reduced the data independently using the \texttt{exoTEDRF} software \citep{Feinstein2023,Radica2023MNRAS.524..835R,Radica2024}.

Data reduction with \texttt{exoTEDRF} consists of three stages.
Stage 1 is detector-level processing which performs bias and dark subtraction, 1/f noise correction, and cosmic ray rejection.
Stage 2 is the spectroscopic processing where flat-fielding, background-subtraction, bad pixel masking, and a principle component analysis (PCA) are performed.
Eigenvalues from the PCA analysis are used to de-trend the data.
The first component of the PCA corresponds to the astrophysical components of the observation and are not removed, so we used only components 2, 3, and 4.
These components correspond to drifts in the position of the trace (components 2 and 3) and the characteristic beating pattern induced by the thermal cycling of the telescope that is common to all NIRISS/SOSS observations (component 4).
Stage 3 of the \texttt{}{exoTEDRF} reduction is where the 1D spectrum is extracted from along the detector with a simple box extraction using \texttt{exoTEDRF}’s built-in optimization to determine the ideal box width (30 pixels in our case). 
This software also has a fourth stage which will perform light curve fitting, but we perform this part ourselves.
Further details of this pipeline and its procedure can be found in the \texttt{exoTEDRF} tutorial\footnote{\url{https://exotedrf.readthedocs.io/en/latest/content/notebooks/tutorial_niriss-soss.html}} and are discussed in \citep{Radica2023MNRAS.524..835R,Radica2024}.

We trim the noisier wavelength regions near the edges of the detector (blue-ward of 0.62~$\mu$m and red-ward of 2.82$\mu$m). We then model the resulting light curves by accounting for planetary transit and  eclipses, phase variations in the planet's thermal emission and reflected light, stellar rotational modulation, and instrumental systematics.
The exoplanet transit and eclipse models are  calculated with \texttt{batman} \citep{batman} and an optimized set of parameters is found using \texttt{scipy.optimize}. Details about the transit model can be found in Appendix \ref{section:appendix-transit-fitting}.
We subtract the best-fitting model from each light curve to obtain the residual time series  (Figure \ref{fig:transit-models}) which we use to investigate stellar granulation and oscillations in the presence of observational noise and any remaining instrumental systematics.

\subsubsection{Solar Observations}
To provide a solar reference for interpreting the granulation and oscillation signatures measured in LTT-9779 and WASP-121, we also analyze Sun-as-a-star observations from the VIRGO experiment on board the Solar and Heliospheric Observatory (SOHO).
In particular, we use VIRGO/PMO6 Total Solar Irradiance (TSI) measurements \citep{Frohlich1995} obtained at a cadence of 1~min between February 1996 and October 2018. 
We divide the more than 8000 days of observations into 48-hour segments, calculate the power spectral density of each segment, and average the resulting spectra. This procedure reduces the stochastic scatter in the solar power spectrum while using individual time series with durations comparable to the JWST observations.

\subsection{Power Spectrum Analysis}
\label{subsection:methods-compute-power-spectra}

We calculate the discrete fast Fourier transform of the de-trended light curves with \texttt{numpy.fft.fft} and convert that Power Spectrum (PS) into a Power Spectral Density (PSD):
\begin{equation}
    \mathrm{PSD}(\nu)=2\Delta t\frac{|\mathrm{PS}(\nu)|^2}{\mathrm{N}},
\end{equation}
where $\nu$ is frequency, $\Delta t$ is the time cadence between exposures, and N is the number of exposures. 
For the chromatic analysis, we chose 6 wavelength bins and summed the light curve along wavelength for each bin.
Bin edges were defined by eye such that there was similar scatter in the residual (de-trended) time series for each bin.

We model the power spectrum, $P(\nu)$, as a sum of Harvey-like profiles \citep{Harvey1985, Stahn2010}, a white-noise term, and a Gaussian envelope representing the  power excess due to   solar-like oscillations:
\begin{equation}\label{eqn:power-spectrum-model}
    P(\nu) = P' + P_{\mathrm{osc}}(\nu) + \sum_{i=1}^{N} \frac{A_i}{1 + (B_i \nu)^{C_i}},
\end{equation}
where $\nu$ is the linear frequency, $P'$ is a frequency-independent white-noise component, and $N_{\rm components}$ is the number of Harvey-like granulation components, taken to be one or two in this work.
For each component, $A_i$ sets the lower-frequency plateau of the power spectral density, $B_i$ determines the turnover frequency and is related to the characteristic timescale, and $C_i$ controls the high-frequency slope of the profile.
The oscillation component, $P_{\mathrm{osc}}$, is modeled as a Gaussian envelope: 
\begin{equation}
    P_{\mathrm{osc}}(\nu) = H_{\mathrm{osc}} \exp \left[ -\frac{(\nu - \nu_{\mathrm{max}})^2}{2\sigma_{\mathrm{env}}^2} \right],
\end{equation}
where $H_{\mathrm{osc}}$ is the peak power spectral density of the envelope,$\nu_{\mathrm{max}}$ is its central frequency, and $\sigma_{\mathrm{env}}$ is the envelope width.

We relate model parameters $A_i$ and $B_i$ to their physical counterparts in the time domain, \sigmagran~(RMS) and \taugran~(s), with 
$$\tau_{\mathrm{gran},i} = \frac{B_i}{2\pi}, \mathrm{and}$$
$$\sigma_{\mathrm{gran},i} = \sqrt{\frac{A_i}{\xi_{C_i} B_i}},$$
where $\xi_{C_i}$ is a normalization constant which satisfies
$$\int_0^\infty \frac{\xi_{C_i} B_i}{1 + (B_i \nu)^{C_i}} \, d\nu = 1.$$
For $C_i = 2$ or $C_i = 4$, analytic solutions for $\xi_{C_i}$ work out to
$$\xi_{2} = \frac{2}{\pi},$$ 
and
$$\xi_{4} = \frac{2\sqrt{2}}{\pi},$$ respectively \citep{Kallinger2014A&A...570A..41K}.
For other values of $C_i$, this integral must be evaluated numerically.

The exponent $C$ controls the high-frequency slope of the Harvey-like profile: $C=2$ is routinely applied in the literature and corresponds to an instantaneously appearing and gradually decaying pattern of granulation variability, while $C=4$ is related to a symmetric rise and decay which exhibits a steeper slope in the power spectrum.
Based on a statistical comparison of one- versus two-component fits and a variety of treatments of C (see Appendix \ref{section:appendix-model-selection}), we focus our results and discussion on three models. Summarized in Table~\ref{tab:model-details}: a one-component model with $C=2$ (hereafter \modelone), a one-component model with $C=4$ (\modeltwo), and a two-component model with $C\in\{2,4\}$ (hereafter \modelthree).
Although \modelthree~includes two Harvey-like components, we do not interpret the second, longer-timescale component as a secure detection of an additional granulation signal. Instead, we treat it as a contribution that accounts for excess low-frequency power that may arise from residual instrumental systematics, imperfect subtraction of planetary signals, or other stellar variability.

\begin{table}[h]
    \centering
    \begin{tabular}{lcc}
        Model                       &   N$_{\mathrm{components}}$                &   $C$        \\
        \hline
        \modelone & 1 & $2$  \\
        \modeltwo & 1 & $4$  \\
        \modelthree & 2 & High-frequency $=2$, \\ & & Low-frequency $=4$ \\
         \hline
    \end{tabular}
    \caption{Details of the three granulation profile models described in this work. For \modelthree, only the high-frequency component is associated with granulation.}
    \label{tab:model-details}
\end{table}

\subsection{MCMC Setup}
\label{subsection:methods-MCMC}
We infer the power spectrum parameters using Bayesian Inference.
We use the \citet{GoodmanWeare2010} affine-invariant Markov Chain Monte Carlo (MCMC) with the \texttt{emcee} \citep{Foreman-Mackey2012} ensemble sampler to compute the marginalized probability functions.
In the minimization we use the likelihood function in the form of the exponential distribution, which is appropriate for power spectra fitting \citep[see equations 9 and 10 of ][]{Larsen2025}.
Models are run with 50 walkers, 60000 steps, and 30000 burn-in steps.
As shown in Table \ref{tab:priors} all parameters have uniform priors with wide bounds with the exception of \numax~and \sigmaenv~which we restrict to reasonable ranges to avoid spurious solutions far from their expected values.

\begin{table}[hb]
    \centering
    \begin{tabular}{lcc}
        Parameter                       &   LTT-9779                    &   WASP-121        \\
        \hline
        $\log\sigma_{\mathrm{noise}}$   & $\mathcal{U}(4,\ 7)$      & $\mathcal{U}(4,\ 7)$ \\
        $\log A_1$                      & $\mathcal{U}(5,\ 8)$      & $\mathcal{U}(5,\ 8)$ \\
        $\log B_1$                      & $\mathcal{U}(2.5,\ 4)$      & $\mathcal{U}(2.5,\ 4)$ \\
        $\log A_2$                      & $\mathcal{U}(4,\ 8.0)$      & $\mathcal{U}(4,\ 8)$ \\
        $\log B_2$                      & $\mathcal{U}(B_1,\ 4.5)$      & $\mathcal{U}(B_1,\ 4.5)$ \\
        $H_{\mathrm{osc}}$              & $\mathcal{U}(4,\ 7)$      & $\mathcal{U}(4,\ 7)$ \\
        $\nu_{\mathrm{max}}$            & $\mathcal{U}(2800,\ 4200)$    & $\mathcal{U}(1300,\ 2100)$ \\
        $\sigma_{\mathrm{env}}$         & $\mathcal{U}(100,\ 500)$      & $\mathcal{U}(10,\ 200)$ \\
         \hline
    \end{tabular}
    \caption{Priors used in the power spectra analysis.}
    \label{tab:priors}
\end{table}

\section{Results} \label{section:results}

\subsection{Solar Power Spectrum}
\begin{figure}[ht]
    \centering
    \includegraphics[width=0.98\linewidth]{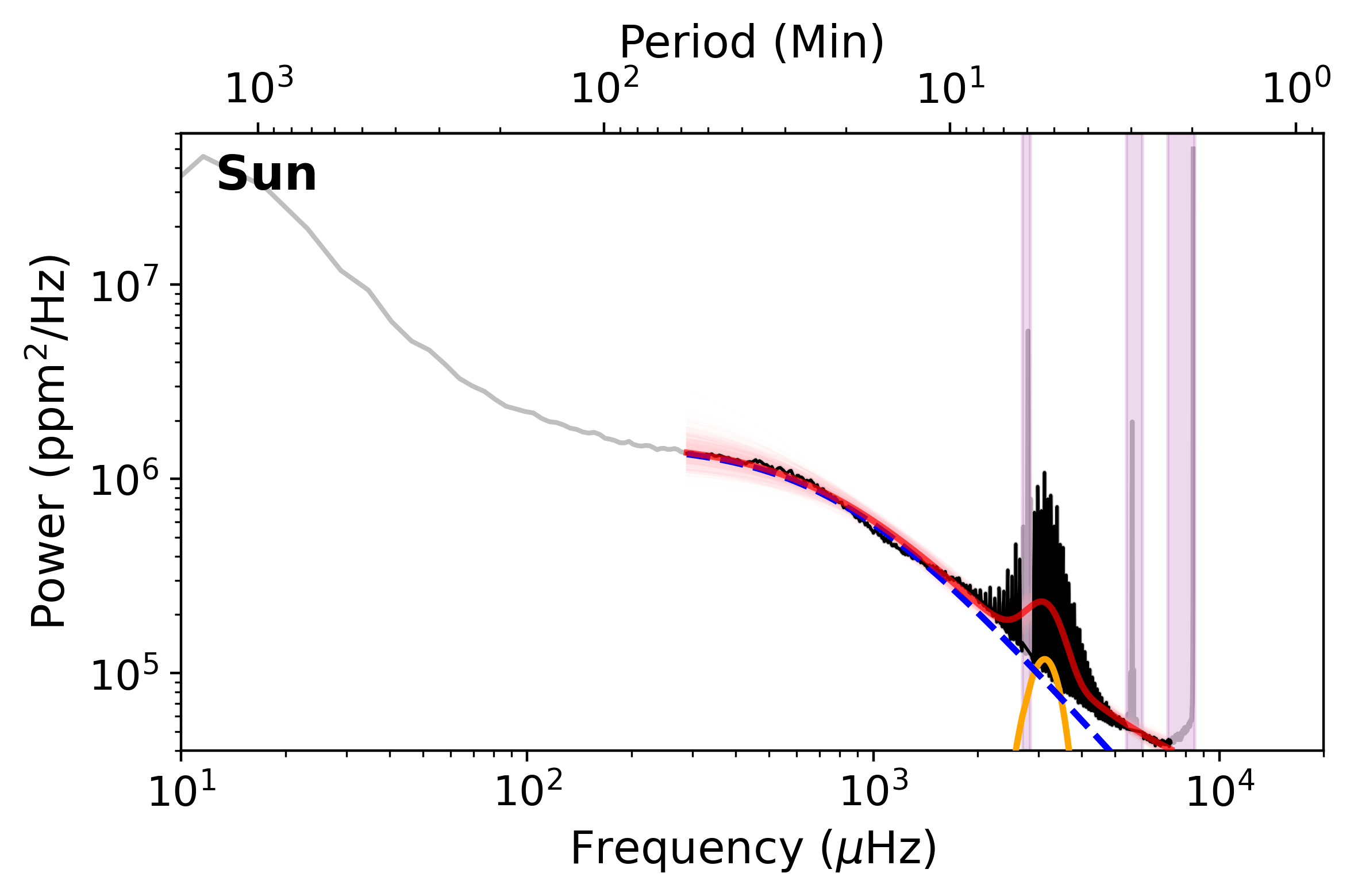}
    \caption{TSI Power spectrum data with a single-component granulation model.
    The background (grey) data is the un-trimmed power spectrum with the systematics-removed power in black.
    Vertical purple bars mark the frequencies we mask because they correspond to the time series sampling interval and its harmonics.
    The red line denotes the best-fit (median) model (bold) with random models drawn from the posterior samples. Blue and orange dashed curves show the separate granulation (blue) and oscillation (orange) components. While the rising power at lower frequencies is related to other sources of stellar activity, the apparent ``plateau'' around 1000 minutes is not a resolved component but an artifact arising from calculating the power spectrum from a finite time series (see Appendix \ref{section:appendix-model-selection}).}
    \label{fig:broadband-model-results-TSI}
\end{figure}

As a benchmark for our analysis, we fit the solar power spectral density with \modelone, which includes a single Harvey-like granulation component with $C=2$, a Gaussian oscillation envelope, and a white-noise term. 
Before fitting, we mask a prominent instrumental peak and two associated harmonics which are caused by the time series sampling frequency. The fitted spectrum is shown in Fig.~\ref{fig:broadband-model-results-TSI}, and the inferred parameters are listed in Table~\ref{tab:broadband-granulation-results}. We adopt this model as the solar reference without performing the additional model comparisons applied to the JWST targets.

\begin{figure*}[ht]
    \centering
    \includegraphics[width=0.48\linewidth]{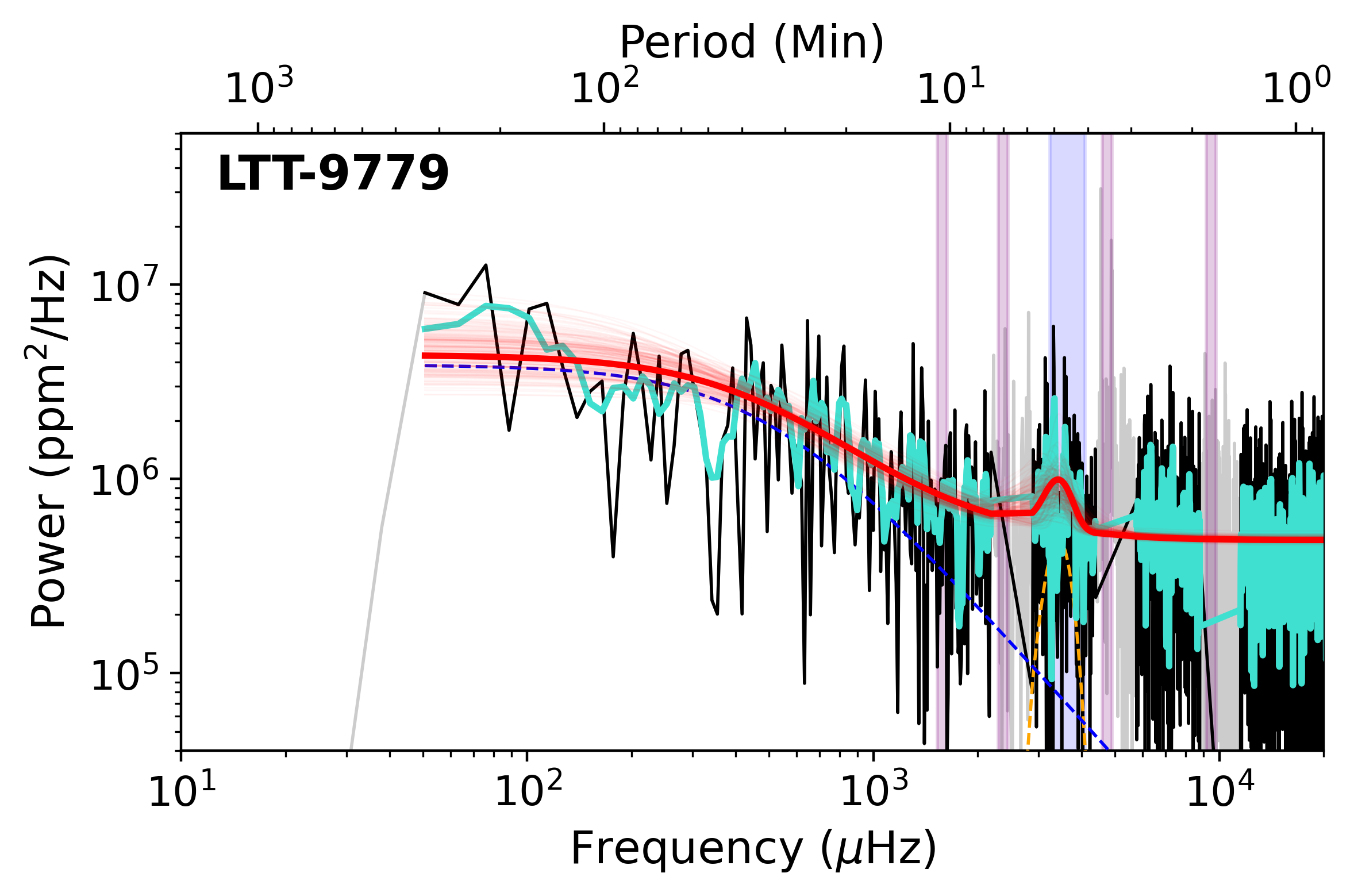}
    \includegraphics[width=0.48\linewidth]{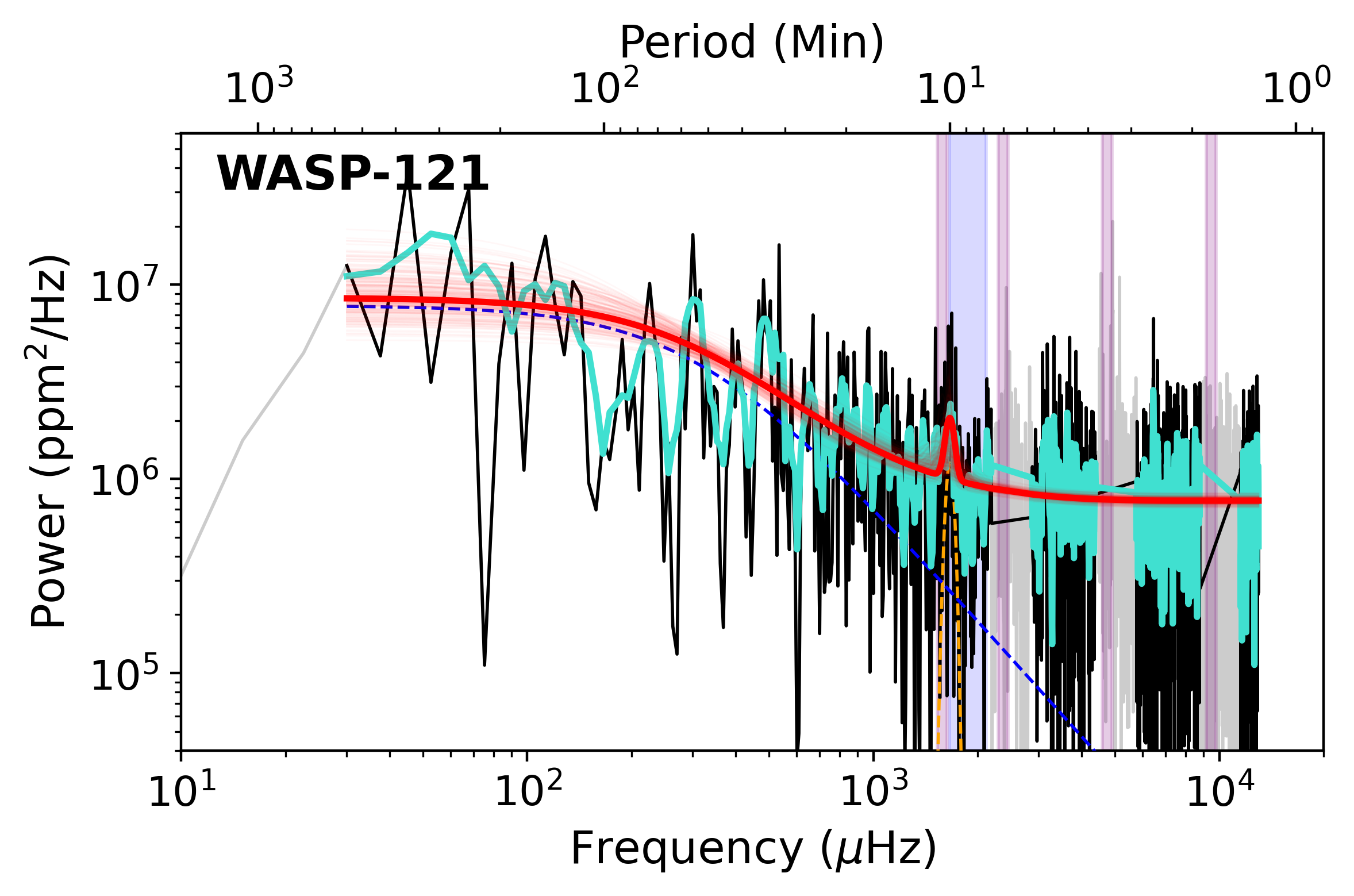} \\
    \includegraphics[width=0.48\linewidth]{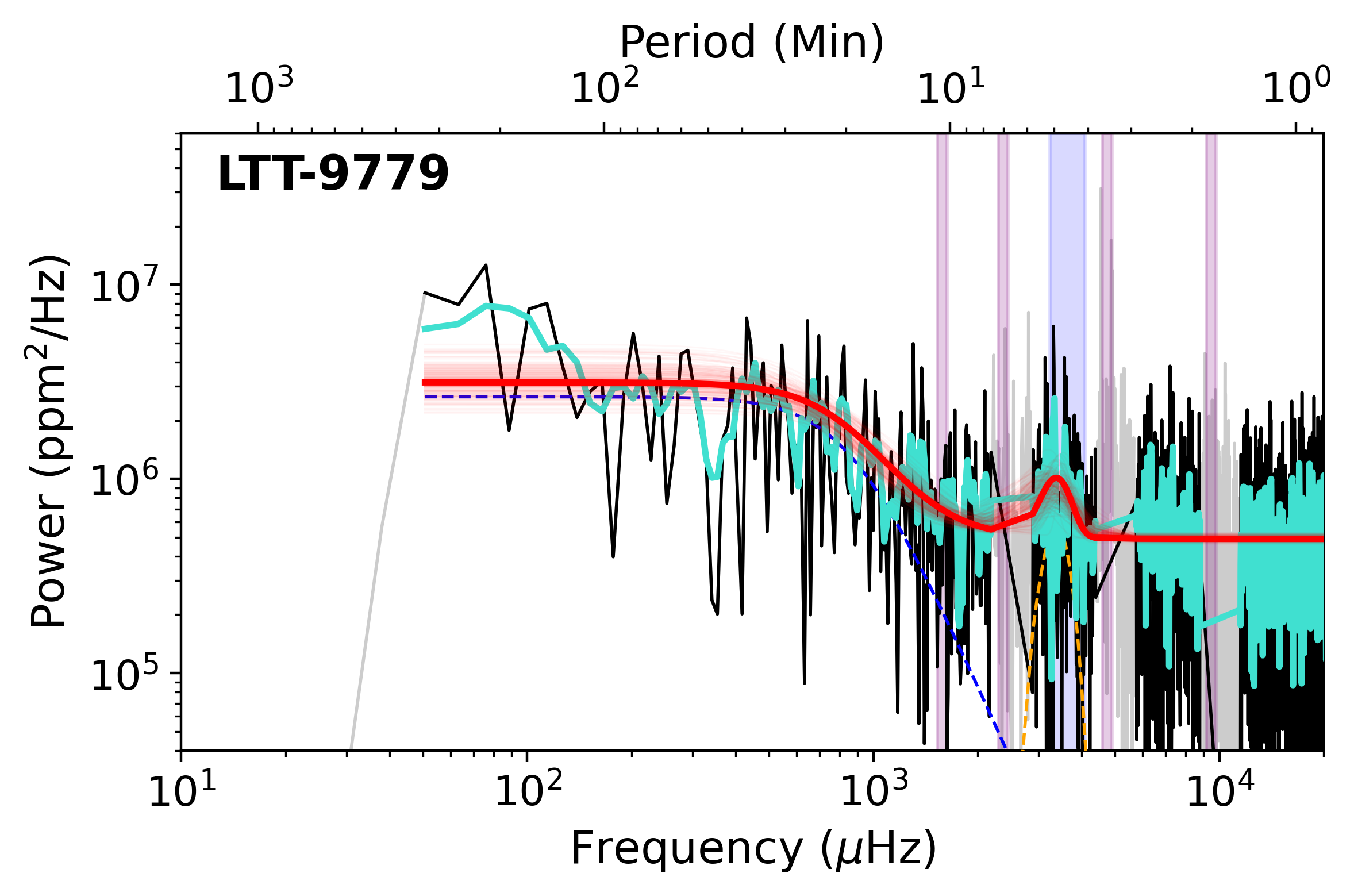}
    \includegraphics[width=0.48\linewidth]{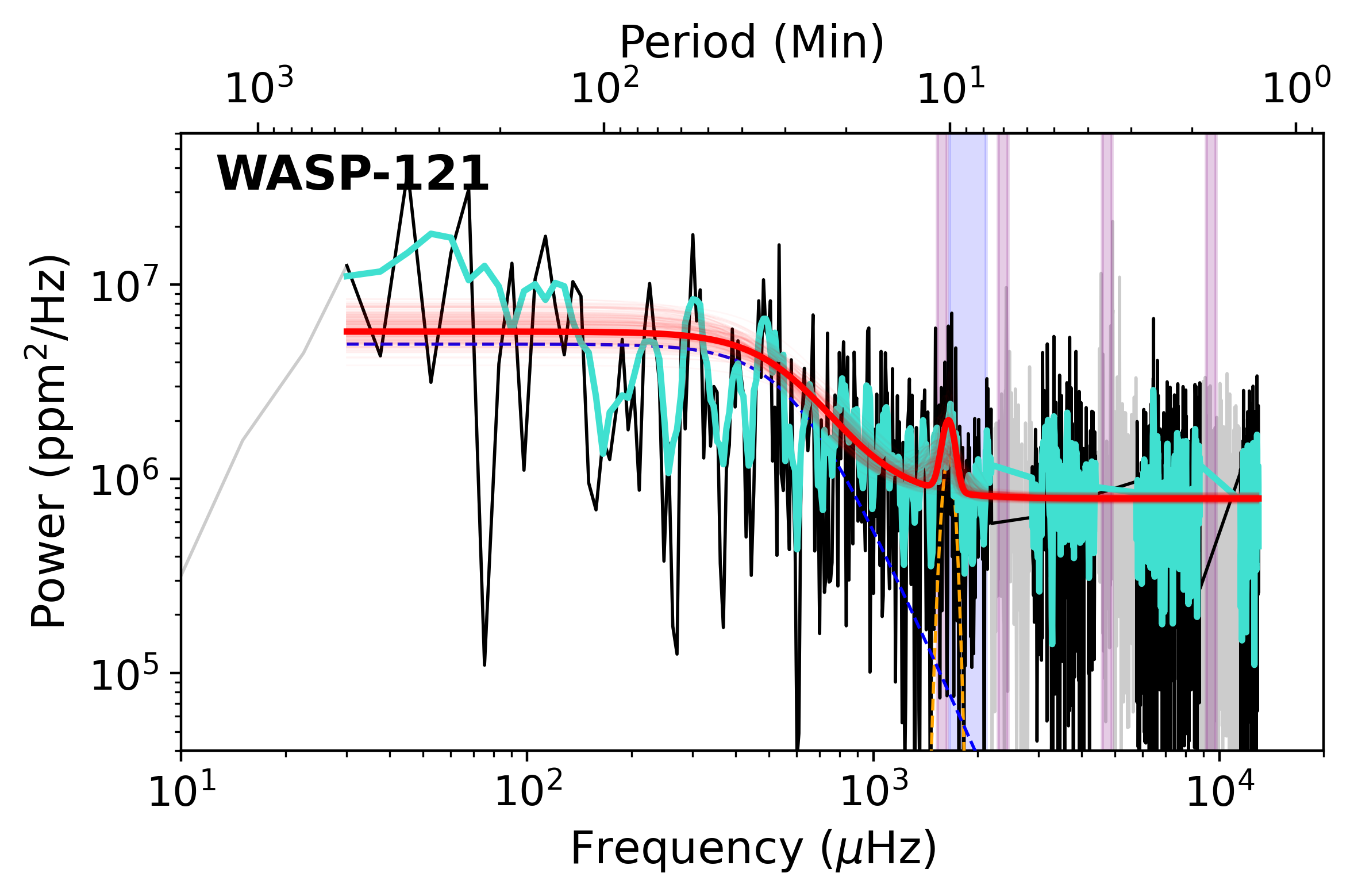} \\
    \includegraphics[width=0.48\linewidth]{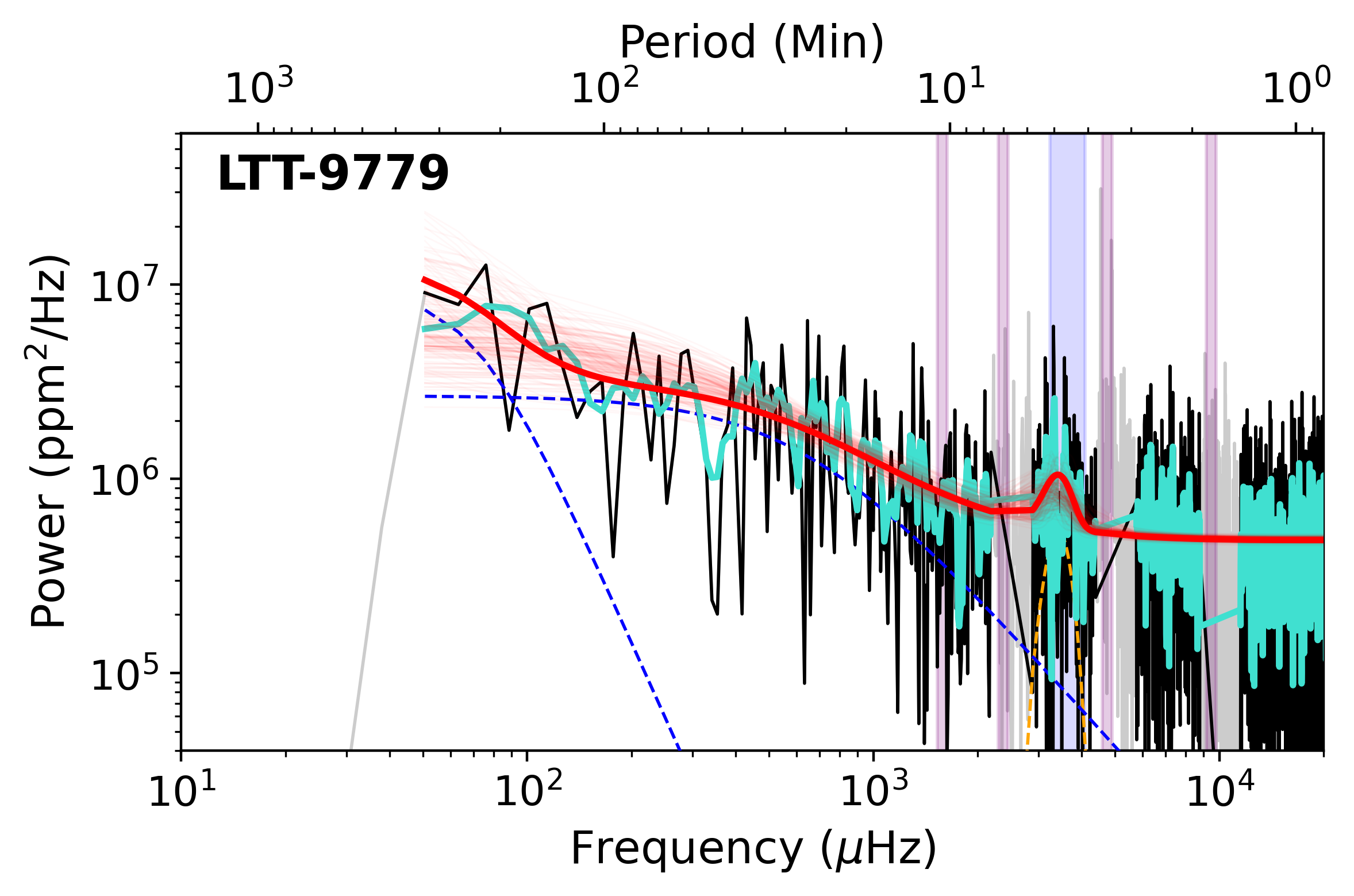}
    \includegraphics[width=0.48\linewidth]{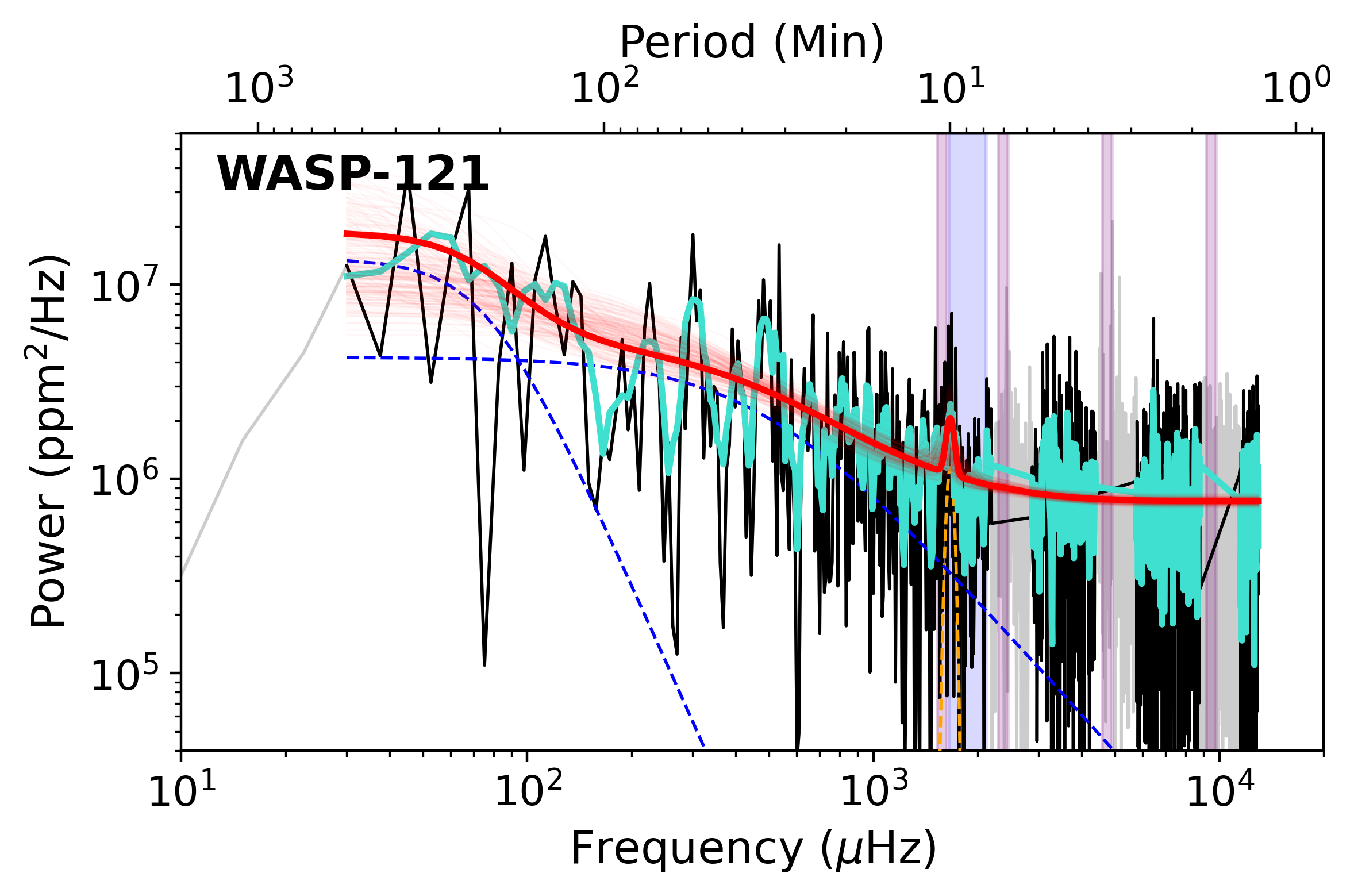} \\
    \caption{Models for the Bandpass-Integrated (BI) time series observations of LTT-9779 (left), and WASP-121 (right) for \modelone~(top), \modeltwo~(middle), and \modelthree~(bottom).
    The background (grey) data is the un-trimmed power spectrum with the systematics-removed power in black.
    Vertical purple bars indicate the instrumental signature and expected harmonics, while the vertical blue bars highlight the predicted range for the oscillation peak, \numax, based on the measured solar oscillations and the stellar parameters of our targets.
    The red lines denote the median parameter model (bold) with random models drawn from the posterior samples.
    A smoothed PSD is plotted in turquoise.
    Blue dashed curves represent the individual maximum-likelihood Harvey Profile components.}
    \label{fig:broadband-model-results}
\end{figure*}

\begin{table}[ht]
    \centering
    \begin{tabular}{lccc}
        \hline
        Parameter                       & Sun                     & LTT-9779      & WASP-121 \\ 
        \hline
        \modelone & & & \\
        $\sigma_{\rm gran}$ [ppm]       & $43.6^{+1.5}_{-1.4}$    & $55.3^{+3.8}_{-3.5}$  & $62.8^{+3.9}_{-3.6}$\\
        $\tau_{\rm gran}$ [min]         & $3.4^{+0.5}_{-0.4}$     & $5.6^{+1.5}_{-1.1}$   & $8.5^{+2.0}_{-1.5}$\\
        $\nu_{\rm max}~[\mu \rm Hz]$    & $3129^{+79}_{-76}$      & $3377^{+164}_{-191}$  & $1647^{+51}_{-67}$ \\
        $\sigma_{\rm env}~[\mu \rm Hz]$ & $372^{+99}_{82}$        & $395^{+174}_{-136}$   & $83^{+80}_{-42}$\\
        \hline
        \modeltwo & & & \\
        $\sigma_{\rm gran}$ [ppm]       &                         & $50.0^{+3.4}_{-3.2}$  & $57.2^{+3.6}_{-3.3}$ \\
        $\tau_{\rm gran}$ [min]         &                         & $3.2\pm0.4$           & $4.6\pm0.5$\\
        $\nu_{\rm max}~[\mu \rm Hz]$    &                         & $3325^{+161}_{-197}$  & $1633^{+48}_{-61}$\\
        $\sigma_{\rm env}~[\mu \rm Hz]$ &                         & $466\pm156$           & $109^{+81}_{-48}$\\
        \hline
        \modelthree & & & \\
        $\sigma_1$ [ppm]                &                         & $53.8^{+4.1}_{-4.0}$   & $59.3\pm4.3$ \\
        $\tau_1$ [min]                  &                         & $4.8^{+1.5}_{-1.3}$    & $6.3^{+2.1}_{-1.8}$ \\
        $\sigma_2$ [ppm]                &                         & $14.0^{+15.3}_{-11.9}$  & $26.6^{+11.5}_{-18.5}$ \\
        $\tau_2$ [min]                  &                         & $33.0^{+54.2}_{-20.0}$ & $38.1^{+48.5}_{-16.0}$ \\
        $\nu_{\rm max}~[\mu \rm Hz]$    &                         & $3382^{+166}_{-191}$   & $1649^{+55}_{-69}$\\
        $\sigma_{\rm env}~[\mu \rm Hz]$ &                         & $387^{+179}_{-137}$    & $82^{+82}_{-44}$ \\
        \hline
        \hline
        Predicted & & & \\
        $\sigma_{\rm gran}$ [ppm]       &                         & $42\pm2$                & $57\pm5$\\
        $\tau_{\rm gran}$ [min]         &                         & $3.2\pm0.4$             & $6\pm1$\\
        \hline
        GP & & & \\
        $\sigma_{\rm gran}$ [ppm]       &                         & $66\pm4$                & $59\pm4$/$67\pm5$\\
        $\tau_{\rm gran}$ [min]         &                         & $3.3^{+0.5}_{-0.3}$     & $4.8\pm0.6$/$5.6^{+1.1}_{-0.8}$\\
        \hline
    \end{tabular}
    \caption{Median values with 1-$\sigma$ uncertainties calculated from the sampled posterior distributions for \modelone~(top), \modeltwo~(upper middle), and \modelthree~(middle). \modelone~and \modelthree~are the preferred models for LTT-9779 and WASP-121, respectively. \modeltwo~is generally the \textit{least} likely model but is included as an analog to the Gaussian Process (GP) kernel used by \citet{Coulombe2025,Splinter2025}. Also included are the scaling relation predictions (middle lower) and literature GP measurements (bottom) which have already been converted from their reported values following the discussion in Sect. \ref{section:discussion:gp-conversions}.}
    \label{tab:broadband-granulation-results}
\end{table}

\begin{figure*}[ht]
    \centering
    \includegraphics[width=0.48\linewidth]{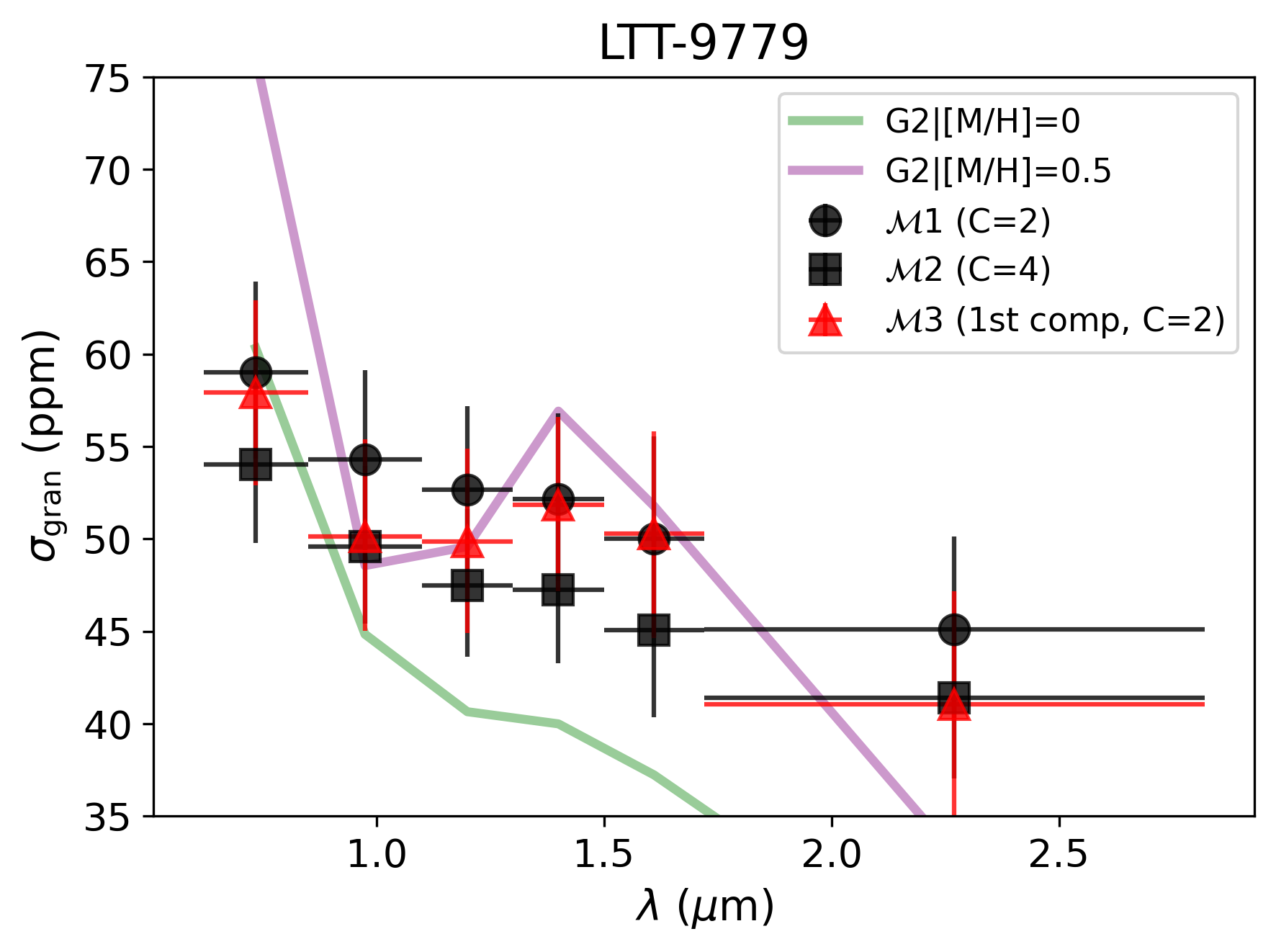}
    \includegraphics[width=0.48\linewidth]{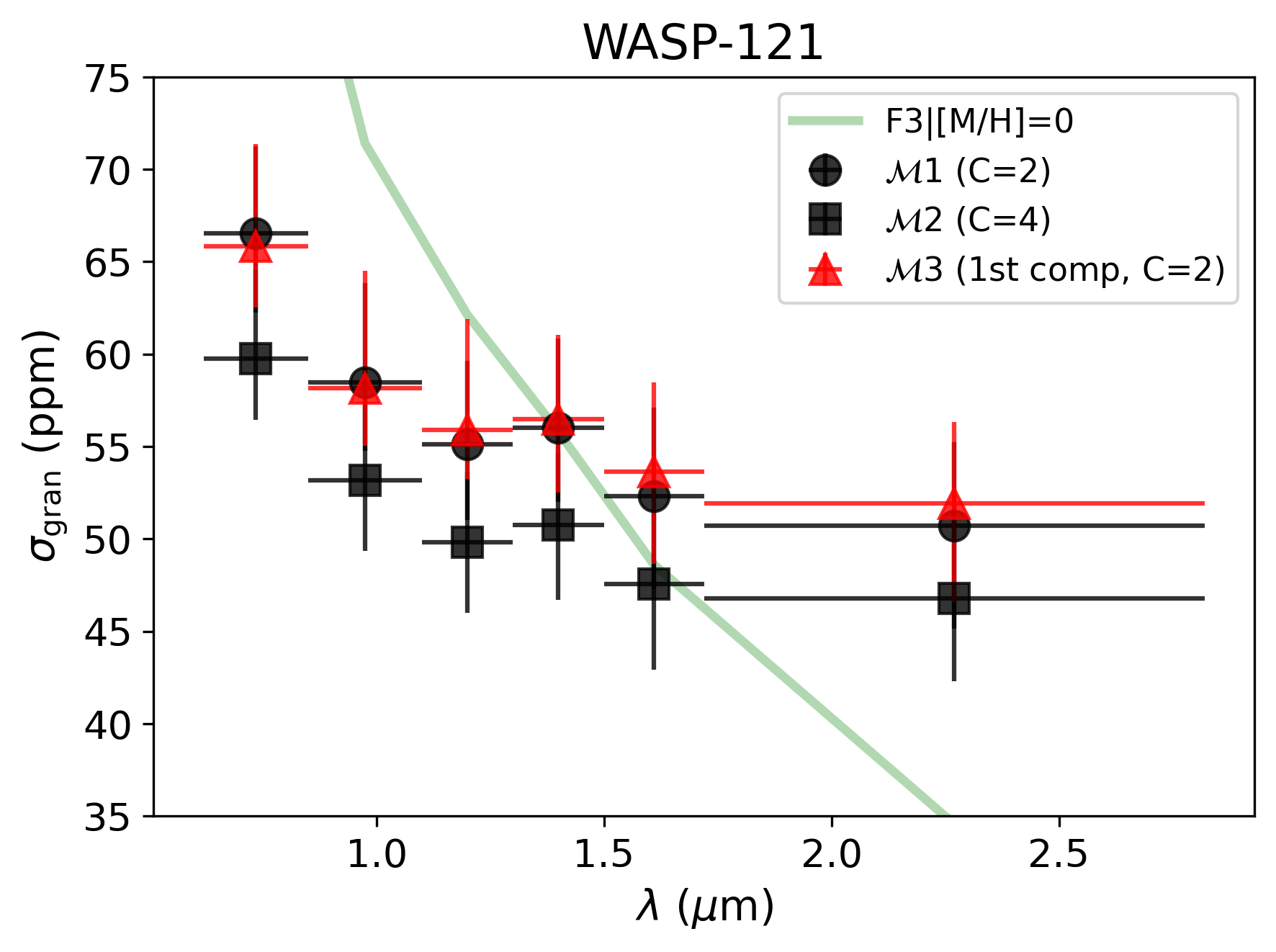}
    \caption{LTT-9779 (left) and WASP-121 (right) chromatic amplitudes measured for each of the three  models. Green and purple curves are results from MURaM simulations for G2 and F3 stars with [M/H]=0 (green) and [M/H]=0.5 dex (purple; only available for G2). Simulations are only shown for by-eye comparison, a quantitative comparison can be done in future work as more simulations become available.}
    \label{fig:chromatic-amplitude-results}
\end{figure*}

\begin{table}[ht]
\centering
    \begin{tabular}{ccccc}
    \hline
    $\lambda$ ($\mu$m)	&	$\Delta\lambda$ ($\mu$m)	&	$\sigma_1$ (ppm)	&	$\sigma_2$ (ppm)	&	$\sigma_{\rm total}$ (ppm)	\\
    \hline
    \modelone &         &	                          &		&	\\
    0.735	  &	0.115	&   $59.0^{+4.9}_{-4.7}$	&		&	\\
    0.975	  &	0.125	&	$54.3^{+4.8}_{-4.6}$    &		&	\\
    1.200	  &	0.100	&   $52.6^{+4.6}_{-4.3}$    &		&	\\
    1.400	  &	0.100	&	$52.2^{+4.7}_{-4.5}$    &		&	\\
    1.610	  &	0.110	&	$49.9^{+5.5}_{-5.3}$      &		&	\\
    2.270	  &	0.550	&	$45.2^{+4.9}_{-4.8}$      &		&	\\
    \hline
    \modeltwo	&		&	     &		&	\\
    0.735	&	0.115	&	$54.0^{+4.5}_{-4.3}$    &		&	\\
    0.975	&	0.125	&	$49.6^{+4.4}_{-4.2}$    &		&	\\
    1.200	&	0.100	&	$47.5^{+4.1}_{-3.9}$    &		&	\\
    1.400	&	0.100	&	$47.2^{+4.2}_{-4.0}$    &		&	\\
    1.610	&	0.110	&	$45.1^{+4.9}_{-4.7}$    &		&	\\
    2.270	&	0.550	&	$41.4^{+4.5}_{-4.4}$    &		&	\\
    \hline
    \modelthree &		&		&		&	\\
    0.735	&	0.115	&	$57.9^{+5.1}_{-4.9}$   &   $16.0^{+16.8}_{-13.0}$   & $74.0^{+15.7}_{-11.4}$\\
    0.975	&	0.125	&	$50.1^{+5.3}_{-5.1}$     & $29.3^{+11.7}_{-11.5}$   & $78.9^{+12.1}_{-11.1}$	\\
    1.200	&	0.100	&	$49.8^{+5.0}_{-4.9}$   &   $26.8^{+13.8}_{-14.9}$  &  $76.2^{+13.4}_{-12.9}$	\\
    1.400	&	0.100	&	$51.7^{+4.8}_{-4.7}$   &   $11.0^{+15.0}_{-8.7}$   &  $63.5^{+13.7}_{-8.8}$\\
    1.610	&	0.110	&	$50.3^{+5.6}_{-5.6}$    &  $8.1^{+16.6}_{-6.3}$   &   $60.3^{+13.8}_{-8.5}$\\
    2.270	&	0.550	&	$40.9^{+6.2}_{-6.4}$    &  $27.8^{+14.6}_{-17.4}$ &   $67.8^{+13.8}_{-13.6}$\\
    \hline
    \end{tabular}
\caption{LTT-9779 chromatic granulation amplitudes. $\sigma_{\rm total}$ is the sum of $\sigma_{1}$ and $\sigma_{2}$.}
\label{tab:ltt-9779-chromatic-results}
\end{table}

\begin{table}
\centering
    \begin{tabular}{ccccc}
    \hline
    $\lambda$ ($\mu$m)	&	$\Delta\lambda$ ($\mu$m)	&	$\sigma_1$ (ppm)	&	$\sigma_2$ (ppm)	&	$\sigma_{\rm total}$ (ppm)	\\
    \hline 
    \modelone &		&		&		&		\\
    0.735	&	0.115	&		$66.6^{+4.6}_{-4.5}$   &		&	\\
    0.975	&	0.125	&		$58.5^{+4.4}_{-4.6}$   &		&	\\
    1.200	&	0.100	&		$55.2^{+4.7}_{-4.0}$   &		&	\\
    1.400	&	0.100	&		$56.0^{+4.1}_{-4.8}$   &		&	\\
    1.610	&	0.110	&		$52.3^{+5.2}_{-4.6}$   &		&	\\
    2.270	&	0.550	&		$50.7^{+5.0}_{-4.9}$   &		&	\\
    \hline
    \modeltwo	&		&		   &		&	\\
    0.735	&	0.115	&		 $59.7^{+3.3}_{-4.9}$   &		&	\\
    0.975	&	0.125	&		$53.2^{+4.0}_{-4.1}$   &		&	\\
    1.200	&	0.100	&		  $49.8^{+3.2}_{-4.5}$ &		&	\\
    1.400	&	0.100	&		$50.9^{+4.0}_{-4.0}$&		&	\\
    1.610	&	0.110	&		$47.6^{+4.0}_{-4.6}$    &		&	\\
    2.270	&	0.550	&		$46.9^{+4.4}_{-4.4}$   &		&	\\
    \hline
    \modelthree &		&		&		&	\\
    0.735	&	0.115	&	$65.9^{+7.7}_{-1.2}$    &  $11.5^{+2.5}_{-20.8}$   &  $78.1^{+4.8}_{-16.7}$ \\
    0.975	&	0.125	&	$58.1^{+5.9}_{-3.5}$    &  $14.1^{+4.4}_{-18.7}$   &  $72.1^{+4.6}_{-15.5}$\\
    1.200	&	0.100	&	$56.0^{+5.4}_{-3.3}$    &  $7.3^{+3.4}_{-14.1}$    &  $64.6^{+3.9}_{-13.0}$\\
    1.400	&	0.100	&	$56.5^{+4.7}_{-3.9}$    &  $6.0^{+2.7}_{-12.0}$     & $63.7^{+3.1}_{-12.2}$\\
    1.610	&	0.110	&	$53.7^{+6.6}_{-3.2}$    &  $5.7^{+3.0}_{-11.8}$     & $61.0^{+4.8}_{-11.1}$\\
    2.270	&	0.550	&	$51.9^{+3.5}_{-6.3}$    &  $3.4^{+8.2}_{-3.4}$      & $56.5^{+9.0}_{-3.6}$	\\
    \hline
    \end{tabular}
\caption{WASP-121 chromatic granulation amplitudes. $\sigma_{\rm total}$ is the sum of $\sigma_{1}$ and $\sigma_{2}$.}
\label{tab:wasp-121-chromatic-results}
\end{table}

\subsection{Bandpass-Integrated Stellar Power Spectra} \label{results-broadband}

While most instrumental systematics are removed in the data processing step, residual instrumental signatures appear as strong peaks in the power spectra (see Figure \ref{fig:broadband-model-results}).
Identified by \citet{Albert2023}, these peaks at 3.40 and 3.63 minutes (204 and 218 seconds or 4902 and 4587 $\mu$Hz) are due to slight forces exerted on the instrument when thermostats are cycled on and off. 
For both targets, we mask the primary signature as well as harmonic signatures at 2$\nu_{\rm Inst}$ and $\frac{1}{2}\nu_{\rm Inst}$.
Additional harmonic signatures may be present but are difficult to identify above the noise. 

Table~\ref{tab:broadband-granulation-results} summarizes the parameters inferred from the three models. The Bayesian information criterion favors \modelone~for both JWST targets, although the Akaike information criterion provides little discrimination between \modelone~and \modelthree~for WASP-121 (see Appendix~\ref{section:appendix-model-selection}). Model \modeltwo~has the same granulation-profile shape as the GP kernel used by \citet{Coulombe2025,Splinter2025}, while \modelthree~includes an additional component to account for excess low-frequency power that is not necessarily attributable to granulation. 
Figure \ref{fig:broadband-model-results} shows the fits obtained with each models for LTT-9779 and WASP-121.

The timescale of the second component in \modelthree~is poorly constrained. Following the interpretation outlined in Sect.~\ref{subsection:methods-compute-power-spectra}, we compare only the first, shorter-timescale component with the granulation components of the single-component models.

\subsection{Chromatic Measurements} \label{results-chromatic}
The chromatic granulation amplitude modeling are shown in Figure \ref{fig:chromatic-amplitude-results} and Tables \ref{tab:ltt-9779-chromatic-results} and \ref{tab:wasp-121-chromatic-results}.
For each model, the characteristic timescale or timescales were fixed to the corresponding posterior medians inferred from the bandpass-integrated PSD.
The reported amplitude uncertainties are therefore conditional on these fixed values. Both stars show an overall decrease in granulation amplitude towards longer wavelengths across all three model prescriptions.

For WASP-121,  \modelone~and the first component of \modelthree~yield similar amplitudes, while \modeltwo~gives amplitudes approximately 6~ppm lower and follows a similar wavelength dependence. For LTT-9779, \modeltwo~also yields systematically lower amplitudes than \modelone. The first component of \modelthree~shows a less smooth wavelength dependence, with a modest local enhancement near 1.4--1.5~$\mu$m. However, this structure is not clearly resolved within the reported uncertainties.

\section{Discussion} \label{discussion}

\begin{figure*}
    \centering
    \includegraphics[width=0.95\linewidth]{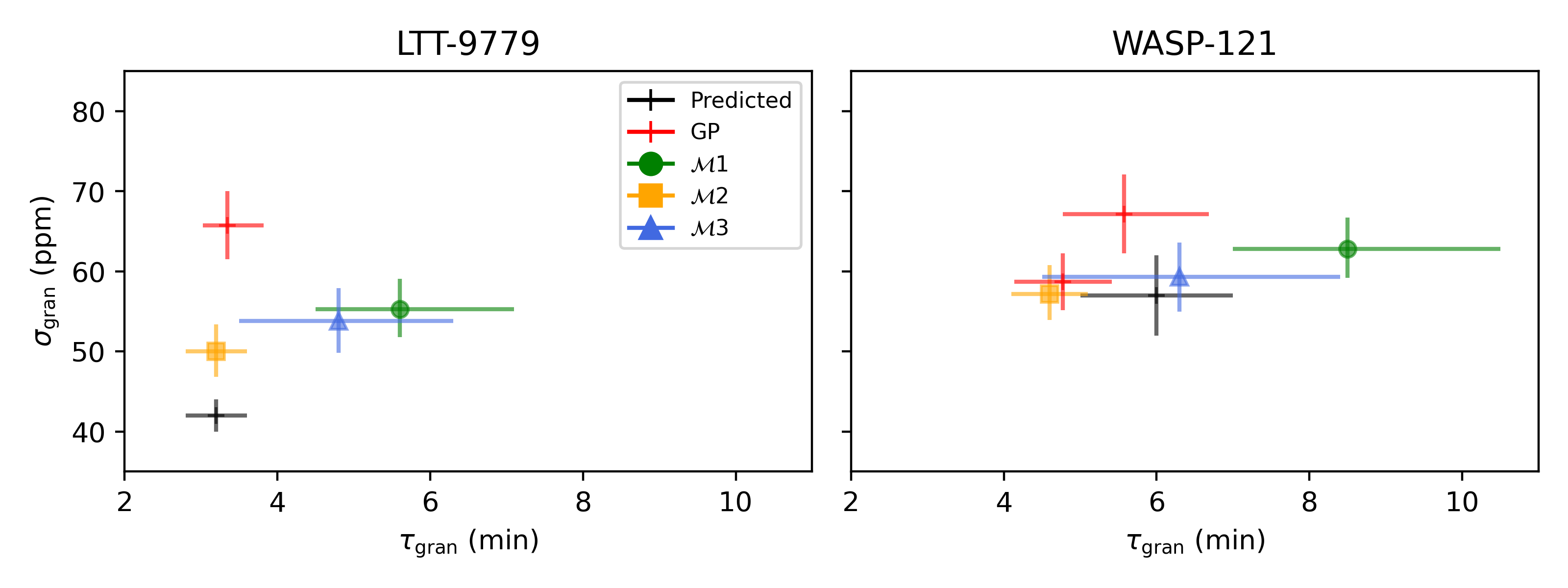}
    \caption{Comparison of our bandpass-integrated measurements to the GP measurements for LTT-9779 (left) and WASP-121 (right). WASP-121 has two GP points because \citet{Splinter2025} report an amplitude and timescale for orders 1 and 2 separately. Also shown are scaling relationship predictions (black) and values derived from this work--\modelone~(green circles), \modeltwo~(orange squares), and \modelthree~(blue triangles). GP measurements have been converted from their reported times and amplitudes into the time and amplitude spaces commensurate with our methods, as described in Sec. \ref{section:discussion:gp-conversions}. }
    \label{fig:comparison-with-gp}
\end{figure*}

\subsection{Oscillation and Granulation Scaling Relations}\label{section:scaling-relations}
We use solar scaling relations to predict the oscillation and granulation parameters of our targets.
The frequency of maximum oscillation power, $\nu_{\mathrm{max}}$, is predicted using the scaling relation of \citet{KjeldsenBedding1995}:
\begin{equation} \label{eqn:nu_max_scaling_relation}
    \frac{\nu_{\mathrm{max}}}{\nu_{\mathrm{max},\odot}} = \left(\frac{M}{M_{\odot}}\right) \left(\frac{R}{R_{\odot}}\right)^{-2} \left(\frac{T_{\mathrm{eff}}}{T_{\mathrm{eff},\odot}}\right)^{-\frac{1}{2}},
\end{equation}
where $M$, $R$, and \Teff are the stellar mass, radius, and effective temperature, respectively, and the subscript $\odot$ denotes solar reference values.
For the characteristic granulation timescale and RMS amplitude, we use the relations given by \citet{Gilliland2011}:
\begin{equation} \label{eqn:tau_gran_scaling_relation}
    \frac{\tau_{\mathrm{gran}}}{\tau_{\mathrm{gran},\odot}} = \left(\frac{M}{M_{\odot}}\right)^{-1} \left(\frac{R}{R_{\odot}}\right)^2 \left(\frac{T_{\mathrm{eff}}}{T_{\mathrm{eff},\odot}}\right)^{1/2},
\end{equation}
and:
\begin{equation} \label{eqn:sigma_gran_scaling_relation}
    \frac{\sigma_{\mathrm{gran}}}{\sigma_{\mathrm{gran},\odot}} = \left(\frac{M}{M_{\odot}}\right)^{-\frac{1}{2}} \left(\frac{R}{R_{\odot}}\right) \left(\frac{T_{\mathrm{eff}}}{T_{\mathrm{eff},\odot}}\right)^{\frac{1}{4}},
\end{equation}
When calculating predictions from the scaling relations, we use the median parameters measured from \modelone~fits of the solar data (also found in Table \ref{tab:broadband-granulation-results}):
$$\sigma_{\odot}=44~\mathrm{ppm, }$$
$$\tau_{\odot}=3.4~\mathrm{min, }$$
and
$$\nu_{\mathrm{max},\odot}=3129~\mu\mathrm{Hz.}$$
From these we predict a granulation amplitude of $42\pm2$ppm with timescale of $3.2\pm0.4$ minutes for LTT-9779.
For WASP-121, we predict an amplitude of $\sigma_{\rm gran}=57\pm5$ppm and timescale of $\tau_{\rm gran}=6\pm1$ minutes.

\subsection{Interpretation of Timescale}

The timescale we measure for the Sun is $3.4^{+0.5}_{-0.4}$ minutes, which agrees with the $\sim3.7$ minute timescale measured in the VIRGO PSD by \citet{Gilliland2011}, but is notably shorter than the 5-6 minute lifetime of large individual solar granules \citep{Title1989,Hirzberger1999}.
On the other hand, \citet{Kallinger2014A&A...570A..41K} fit the VIRGO PSD and measured $\tau_\odot$ to be $~6.3$ minutes.

We emphasize that the timescale inferred from the fit does not necessarily correspond to the full lifetime of a granule.
Instead, it may characterize a particular phase of granular evolution during which the brightness changes most strongly. Fits obtained with different values of $C$ may therefore probe different phases of a granule’s lifetime.

\subsection{Oscillations}

The scaling-relation predictions provide a consistency check on whether the fitted excess power can be attributed to solar-like oscillations.
Using the stellar parameters in Table~\ref{tab:target-parameters} and Equation~\ref{eqn:nu_max_scaling_relation}, we predict $\nu_{\rm max}=3664\pm143~\mu\mathrm{Hz}$ for LTT-9779, compared with our measured value of $3382^{+166}_{-191}~\mu\mathrm{Hz}$.
For WASP-121, the predicted frequency is $1888\pm77~\mu\mathrm{Hz}$, whereas the measured value is $1649^{+55}_{-69}~\mu\mathrm{Hz}$.

The interpretation of these measurements is complicated by narrow peaks of excess power around $2200$--$2900~\mu\mathrm{Hz}$, which may be associated with a subharmonic at half the frequency of the NIRISS instrumental beat signature. Our Gaussian model describes the collective power of the modes of oscillation modes rather than individual modes. Masking potentially instrumental peaks can therefore also remove stellar oscillation power, reducing the information available to constrain the central frequency and width of the oscillation envelope.

For LTT-9779, the measured $\nu_{\rm max}$ is broadly consistent with the scaling-relation prediction, supporting a stellar origin for the fitted excess power.
For WASP-121, the measured frequency is lower than predicted, and possible contamination by instrumental features makes the interpretation less secure. We therefore regard the identification of solar-like oscillations in WASP-121 as tentative.

\subsection{Granulation Parameters in GP and Power Spectrum Analysis} \label{section:discussion:gp-conversions}

Recent investigations have analyzed granulation signatures directly in the light curve using Gaussian Processes (GP) in order to enhance transit modeling \citep{Pereira2019,Barros2020,Cadieux2024,Coulombe2025,Splinter2025,Pass2026}.
Here we compare the granulation amplitudes and characteristic timescales inferred from our analysis with both scaling-relation predictions and published GP-based measurements.
To do this comparison we must first translate the parameters reported by the GP analyses, which used the angular-frequency SHO of the \texttt{celerite} package \citet{celerite}, into the linear frequency space of the Harvey model.
We also attempt to reconcile the fact that the timescale and amplitude measured by a GP are larger than predicted from scaling relations and measured for the Sun \citep[see discussion in ][]{Pass2026}.

\subsubsection{Timescale}
We start by examining the timescales.
Part of the apparent discrepancy comes from a difference between the timescale definitions adopted by  and in our Harvey-like power-spectrum model.
In the GP approach, the exact form of the simple harmonic oscillator (SHO) profile used by \texttt{celerite} is:
\begin{equation}\label{eqn:sho_celerite_exactform}
    P(\omega)_{GP} = \sqrt{\frac{2}{\pi}} \frac{S_0\omega_0^4}{(\omega^2-\omega_0^2)+\omega^2\omega_0^2Q^{-2}},
\end{equation}
which, for $Q=1/\sqrt{2}$, can be simplified to:
\begin{equation}\label{eqn:sho_gp_pomega}
    P(\omega)_{GP} = \sqrt{\frac{2}{\pi}} \frac{S_0}{(1+(\omega/\omega_0)^4)},
\end{equation}
with the timescale defined as $$\tau_{\mathrm{gran,GP}}=\frac{2\pi}{\omega_0}.$$ 
If we compare the denominator of eqn. \ref{eqn:sho_gp_pomega} to the Harvey profile denominator in eqn. \ref{eqn:power-spectrum-model}, we can see that $\omega/\omega_0=B\nu$.
With the definition that linear frequency $\nu$ relates to angular frequency $\omega$ as $\omega=2\pi\nu$, we arrive at:
\begin{equation}
    \tau_{\mathrm{gran}}=\frac{\tau_{\mathrm{gran, GP}}}{2\pi}.
\end{equation}

In the following discussion, we have taken the reported (GP-derived) timescales from \citet{Coulombe2025} and \citet{Splinter2025} and divided them by 2$\pi$, calculating the following timescales: 
$$\tau_{\mathrm{gran,LTT9779}}=3.3^{+0.5}_{-0.3}~\mathrm{min}, $$
$$\tau_{\mathrm{gran,WASP121|O1}}=4.8\pm0.6~\mathrm{min,~and} $$
$$\tau_{\mathrm{gran,WASP121|O2}}=5.6^{+1.1}_{-0.8}~\mathrm{min}. $$

\subsubsection{Amplitude}

There is also a significant difference between the GP-derived granulation signal and both the scaling relation predictions and our results.
We find a resolution by again examining the analytic forms of the simple harmonic oscillator used in our method and the GP method, arriving at the following relation (derivation provided in Appendix \ref{section:appendix-derivation-of-sigmagp}):
\begin{equation}\label{eqn:sigma_conversion}
    \sigma_{\mathrm{gran},GP}=\frac{a_{\mathrm{gran,}GP}}{\sqrt{2}}.
\end{equation}
We use eqn. \ref{eqn:sigma_conversion} to convert the \citet{Coulombe2025} measurement of $93\pm6$ppm to $66\pm4$ppm and the \citet{Splinter2025} measurements of $83\pm5$ppm and $95\pm7$ppm to $59\pm4$ppm and $67\pm5$ppm for order 1 and order 2, respectively.
Scaling relation predictions and converted GP measurements are listed alongside our measurements in Table \ref{tab:broadband-granulation-results}.

\subsection{Wavelength Dependence of Granulation}

The wavelength dependence of the granulation amplitude provides an additional test of the stellar origin of the measured variability. Whereas the bandpass-integrated amplitude describes the overall photometric fluctuations, its spectral dependence probes how the contrast between bright and dark surface structures changes with wavelength. Comparing this dependence with stellar-atmosphere simulations therefore allows us to assess whether the observed variability is consistent with granulation and to explore its sensitivity to stellar properties.

We measure a wavelength-dependent decrease in amplitude for both stars (see Tables \ref{tab:ltt-9779-chromatic-results} and \ref{tab:wasp-121-chromatic-results}).
To interpret these trends, we compare our measurements with the granulation-induced amplitude spectra of \citet{Kostogryz2026}, calculated from three-dimensional magnetohydrodynamic simulations using the MURaM code.
We use an F3 model with $[\mathrm{M/H}]=0$ as a proxy for WASP-121 and G2 models with $[\mathrm{M/H}]=0$ and $+0.5$ as proxies for LTT-9779. These are the closest available MURaM-simulated models to the stellar parameters of our targets. However, their surface gravities and effective temperatures differ from those of the observed stars. Because granulation-induced variability amplitude depends strongly on surface gravity, this mismatch limits the interpretation of differences in absolute amplitude. Instead, we focus on a qualitative comparison of the spectral shapes. A quantitative comparison will require simulations matched to the surface gravity, effective temperature, and metallicity of each target.

For LTT-9779, the predicted simulated amplitudes are significantly lower than what we measure but reproduce the general spectral trend. Notably, the enhanced metallicity model (purple) appears more similar to that of the spectral profile we measure, demonstrating the promise of characterizing stellar atmospheres from spectral granulation profiles.

For WASP-121 we do not have an enhanced metallicity simulation available. The simulated profile shows agreement with the overall level of signal strength albeit with a much steeper slope than what we measure. 

\section{Summary \& Conclusions} \label{conclusions}

We analyzed the bandpass-integrated and wavelength-resolved power spectra of JWST/NIRISS/SOSS observations of LTT-9779 and WASP-121 to characterize their granulation signatures. Our main findings are as follows.

\begin{itemize}
    \item We measured granulation amplitudes and characteristic timescales from bandpass-integrated timeseries of the G7V-type star LTT-9779 and the F6V-type star WASP-121 using Harvey-like power spectrum models. Comparing one- and two-component models as well as different treatments of C, we can see that there is general agreement between model results, with slightly smaller timescales and amplitudes measured for \modeltwo~compared to \modelone~and \modelthree. 
    For \modelthree, we interpret the second longer-timescale component as contribution from excess low-frequency power (Figure \ref{fig:appendixfig-solar-psd}) rather than as a secure detection of an additional granulation component.
    Under this interpretation, we recommend taking the measurements from the first component of \modelthree~as the best approximation of granulation for these observations.
    This gives us a granulation amplitude of $\sigma_{\mathrm{gran}}=53.8^{+4.1}_{-4.0}$ppm with characteristic timescale $\tau_{\mathrm{gran}}=4.8^{+1.5}_{-1.3}$min for LTT-9779 and granulation amplitude of $\sigma_{\mathrm{gran}}=59.3^{+4.3}_{-4.3}$ppm with characteristic timescale $\tau_{\mathrm{gran}}=6.3^{+2.1}_{-1.8}$min for WASP-121. 

    \item We identified an excess-power envelope consistent with solar-like oscillations in LTT-9779, with $\nu_{\rm max}=3382^{+166}_{-191}~\mu\mathrm{Hz}$, broadly consistent with the scaling-relation prediction. For WASP-121, the fitted value, $\nu_{\rm max}=1649^{+55}_{-69}~\mu\mathrm{Hz}$, is lower than predicted, and we regard the oscillation identification as tentative. Potential instrumental features and the masking of affected frequency intervals limit the interpretation of these envelopes.
    
    \item We compared \citet{Coulombe2025} and \citet{Splinter2025} GP measurements with our results and found that in both cases, the GP measures greater amplitudes and longer timescales than we obtain.
    We believe this is not a physical contradiction or methodological inaccuracy, but a confusion driven by inconsistent definitions of timescale and signal amplitude.
    Following the discussion of this point in the text, we scale the GP measured timescales by $\frac{1}{2\pi}$ and their associated amplitudes ($a_{\rm gran}$) by $\frac{1}{\sqrt{2}}$.
    These conversion factors bring the GP-based measurements closer to our results and the scaling relation predictions.
    
    \item Finally, we investigated the wavelength-dependence of granulation amplitude, finding that both stars' granulation signal decreases with wavelength in ways that are consistent with theoretical models for similar stars. From \citet{Kostogryz2026} we expect granulation amplitude to increase for more active and higher metallicity stars, which we will investigate in future work as part of the ``Granulation in Color'' archival grant.
    
\end{itemize}

These results demonstrate the potential of JWST time-series spectrophotometry to characterize the spectral dependence of stellar granulation and inform its treatment in exoplanet light-curve analyses.

\begin{acknowledgements}
    This work is supported by the Volkswagen Foundation.  N.K., V.V., and L.G. acknowledge financial support through German Aerospace Center (DLR) grant ``PLATO Data Center''  \#50OP1902 and European Research Council (ERC)  Synergy Grant  ``WHOLE SUN'' \#810218. L.G. is supported in part by the joint ANR–DFG project BUTTERFLY under the Agence nationale de la recherche (ANR grant 23-CE46-0009-01) and the Deutsche Forschungsgemeinschaft
(DFG grant 530101854).   
    Data presented in this publication makes use of observations taken with the NASA/ESA/CSA JWST from GO \# 1201 (PI D. Lafrenière), accessed through the Mikulski Archive for Space Telescopes (MAST) hosted at the Space Telescope Science Institute.
    We thank Wolfgang Finsterle for providing the calibrated VIRGO/PMO6 observations.
    In addition to the software cited in the text, we gratefully acknowledge use of the open-source codes astropy \citep{astropy}, chromatic, corner \citep{corner}, matplotlib \citep{matplotlib}, numpy \citep{numpy}, and scipy \citep{scipy}.
    This analysis is associated with JWST AR \#11527 (PIs: J. Valenti \& A. Glidden) with support provided by NASA through a grant from the Space Telescope Science Institute, operated by the Association of Universities for Research in Astronomy, Inc. under NASA contract NAS 5-03127.
    WCW is supported by Volkswagen Foundation grant no. 9E126.
    R.L. is funded by the European Union (ERC, THIRSTEE, 101164189) and acknowledges financial support from the Severo Ochoa grant CEX2021-001131-S funded by MCIN/AEI/10.13039/501100011033. Views and opinions expressed are however those of the author(s) only and do not necessarily reflect those of the European Union or the European Research Council. Neither the European Union nor the granting authority can be held responsible for them. 
    I.R. and J.B.-P. acknowledge support from the European Research Council (ERC) under the European Union’s Horizon Europe programme (ERC Advanced Grant SPOTLESS; no. 101140786), from the Spanish grant PID2024-158486OB-C31 funded by MCIU/AEI/10.13039/501100011033 and by “ERDF A way of making Europe”,  from the programme Unidad de Excelencia María de Maeztu CEX2020-001058-M financed by MCIN/AEI/10.13039/501100011033 and by the MaX-CSIC Excellence Award MaX4-SOMMA-ICE, and from the Generalitat de Catalunya/CERCA programme. J.B.-P. also acknowledges financial support from Spanish grant PID2020-120375GB-I00, funded by MCIU/AEI, Consolidación 2022 CNS2022-136050 and PRE2022-101942 funded by MICIU/AEI/10.13039/501100011033 and ESF+.
\end{acknowledgements}

\bibliographystyle{aa}
\bibliography{main}

\appendix
\nolinenumbers

\section{Light Curve Fitting}\label{section:appendix-transit-fitting}

We model the light curves as the sum of a planetary component and a systematic component, following the general form of the models constructed by \citet{Coulombe2025} and \citet{Splinter2025}.
The planetary component contains the transit and eclipse models along with phase curve components for the planetary reflection and emission.
WASP-121 also requires model components for stellar variability and ellipsoidal variations of the planetary radius, an effect which is observable for gas giants but not for smaller (more rigid) planets. 
The systematic portion of the model accounts for visit-long changes to the baseline and ``tilt'' offsets caused by slight movements of JWST's primary mirror \citep{Rigby2023,Alderson2023}.
The LTT-9779 light curve has two of these events, while WASP-121 has a single event.

Taking all of this into account, we assemble the following model for both targets' light curves: 
\begin{equation}
    F(t) = A(t)+S(t),
\end{equation}
where $F(t)$ is the total normalized flux, $A(t)$ is the astrophysical component, and $S(t)$ is the systematics model. 
The specifics of $A(t)$ and $S(t)$ vary between targets, so we start with LTT-9779:
\begin{equation}
    A(t)=F_*(t) + F_p(t),
\end{equation}
where $F_*(t)$ and $F_p(t)$ are the stellar and planetary flux models. 
For LTT-9779 the $>$45-day rotation period of this star is significantly longer than the $\sim1$-day observational window, so we set $F_*(t)=1.0$
and model $F_p(t)$ as:
\begin{equation}
    F_p(t)=\tau(t)\epsilon(t)[1+f_{\mathrm{p}}(\Phi(t))],
\end{equation}
where $\tau(t)$ is the transit model, $\epsilon(t)$ is the eclipse model, $f_{\mathrm p}$ is the average depth between the two eclipses, and $\Phi(t)$ is the thermal phase curve.
We model $\Phi(t)$ as a 2nd order (i.e., up to $n=2$) Fourier series:
\begin{equation}
    \Phi(t) = 1 + \sum_{n=1}^{2} \left[ C_n \left( \cos(n\theta(t)) - 1 \right) + D_n \sin(n\theta(t)) \right],
\end{equation}
where $C_n$ and $D_n$ are coefficients for $n=1$ and $n=2$ (4 coefficients total), and $\theta(t)$ is the orbital phase, defined as:
\begin{equation}
    \theta(t) = \frac{2\pi (t - T_{0,\mathrm{ecl}})}{P_{\mathrm{ecl}}},
\end{equation}
with eclipse midpoint $T_{0,\mathrm{ecl}}$ and orbital period calculated from the difference in eclipse midpoints, $P_{\mathrm{ecl}}$.

For the systematics model:
\begin{equation}
    S(t) = B(t) + J_1(t) + J_2(t),
\end{equation}
where $B(t)$ is the baseline and $J_{1,2}$ represents the first and second tilt events. 
The out-of-transit baseline is modeled with a 3rd order polynomial:
\begin{equation}
    B(t) = a_0 + a_1 t + a_2 t^2 + a_3 t^3,
\end{equation}
where $a_n$ are free coefficients.
The tilt events are modeled as:
\begin{equation}
    J_{1,2} = j_{1,2} \, H(t - T_{\mathrm{tilt},1,2}),
\end{equation}
where $j_{1,2}$ are the tilt amplitudes, $T_{\mathrm{tilt},1,2}$ represents the time of the tilt events and $H(t - T_{\mathrm{tilt},1,2})$ is the Heaviside function that applies a flux offset to the baseline for all $t > T_{\mathrm{tilt},1,2}$.

For WASP-121 the model looks very similar, with a few key differences. 
Our astrophysical model must now account for ellipsoidal variations of the planet's radius (relevant for gas giants like WASP-121 but not smaller more rigid planets like LTT-9779) as well as stellar rotation:
\begin{equation}
    A(t) = F_*(t) + F_p(t),
\end{equation}
with 
\begin{equation}
    F_*(t) = 1 + A_*\sin(\frac{2\pi}{P_*}(t-t_0)-\phi_*),
\end{equation}
with rotation amplitude $A_*$, stellar rotation period $P_*$ \citep[fixed at 1.13 days; ][]{Sing2024}, mid-transit time $t_0$, and stellar phase offset $\phi_*$.
The planetary contribution is the combined effect of a transit, eclipses, ellipsoidal variation, and thermal phase:
\begin{equation}
    F_p(t) = \tau(t) \epsilon(t) \left[ 1 + f_p \Phi(t) \Omega(t) \right].
\end{equation}
$\Phi(t)$ is a second-order Fourier series and the ellipsoidal variations function $\Omega(t)$ is defined as:
\begin{equation}
    \Omega(t) = \sqrt{ \sin^2 i \left[ \left( \frac{R_{\mathrm{sub}}}{R_p} \right)^2 \sin^2 \theta(t) - \cos^2 \theta(t) \right] + \left( \frac{R_{\mathrm{sub}}}{R_p} \right)^2 \cos^2 i },
\end{equation}
where $i$ is the orbital inclination and $R_{\mathrm{sub}}/R_p$ is the ratio of the planetary radius at the substellar point (the line connecting the star and planet) to the planetary radius along the polar axis. 

Finally, the systematics and baseline model for WASP-121 is very similar to that used for LTT-9779:
\begin{equation}
    S(t) = B(t)+J(t),
\end{equation}
where B(t) is a linear function with a slope and offset, and $J(t)$ is the model for a single tilt event.

\begin{figure*}
    \centering
    \includegraphics[width=0.95\linewidth]{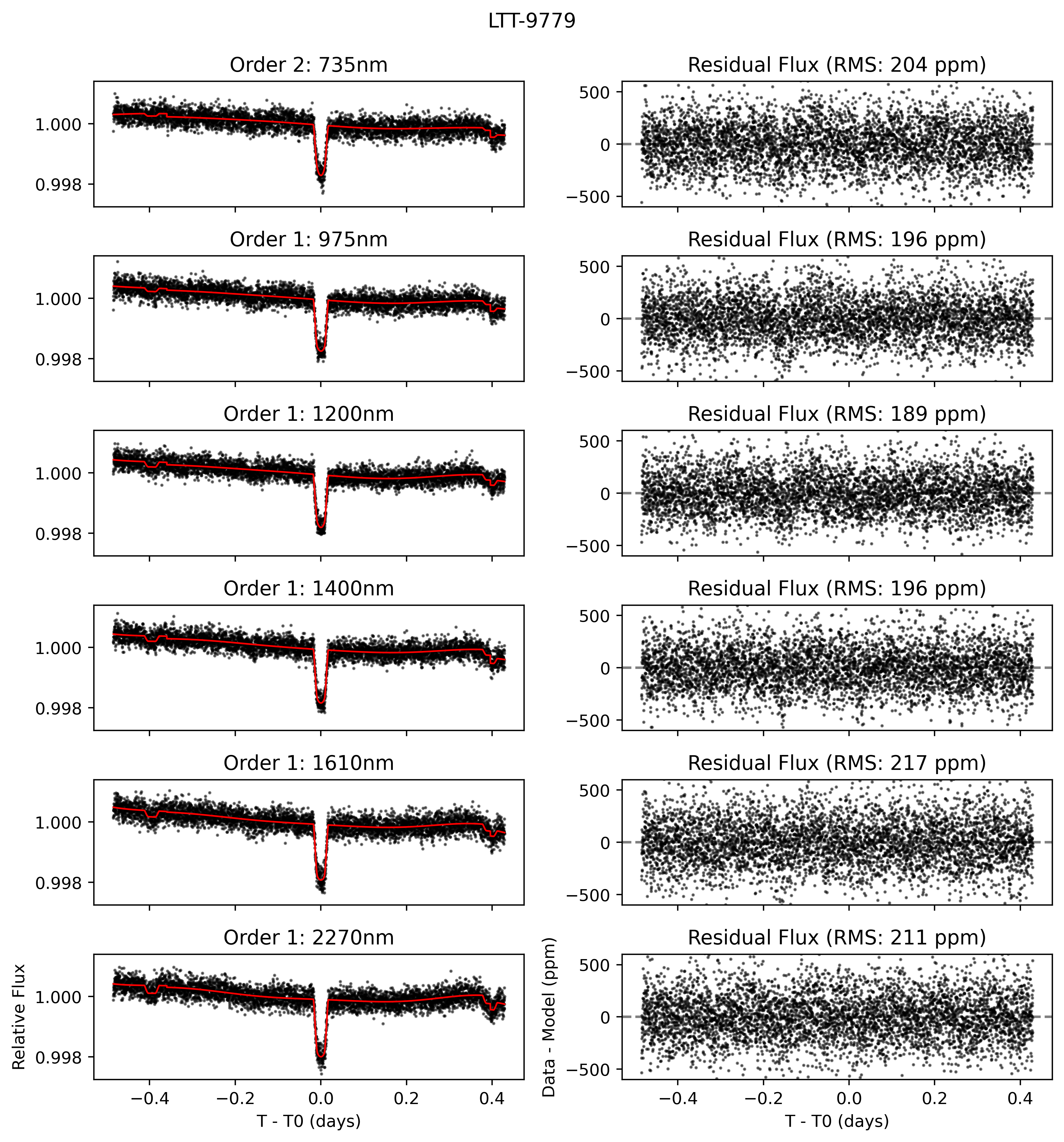}
    \caption{Transit fits and de-trended light curves for LTT-9779.}
    \label{fig:light-curves-ltt}
\end{figure*}

\begin{figure*}
    \centering
    \includegraphics[width=0.95\linewidth]{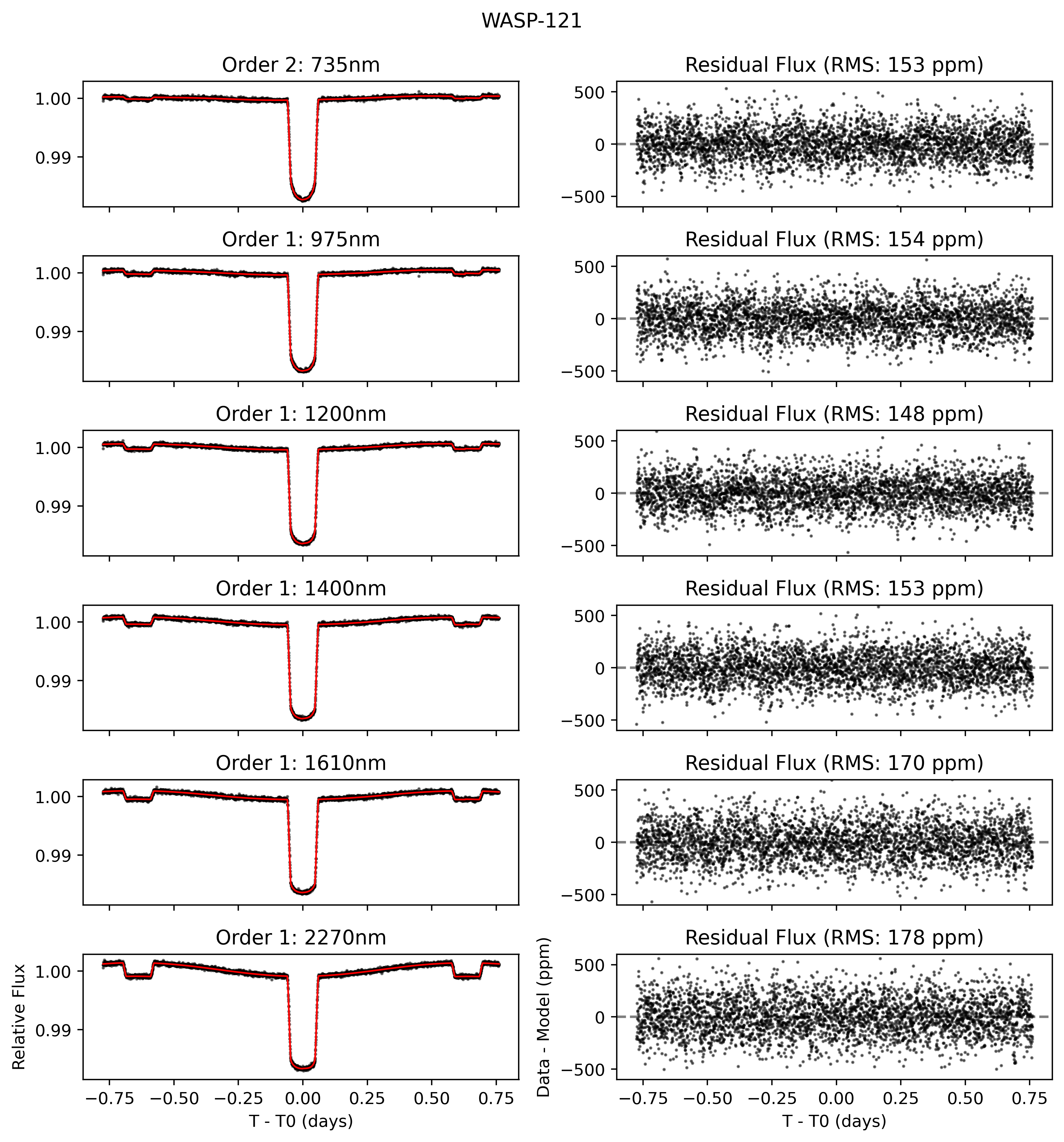}
    \caption{Transit fits and de-trended light curves for WASP-121.}
    \label{fig:light-curves-wasp}
\end{figure*}

In the transit model, we fix eccentricity to zero and fit for the transit midpoint $T_{0,\mathrm{tra}}$, planet radius $R_p/R_*$, scaled semi-major axis $a/R_*$, inclination $i$, and the quadratic limb darkening coefficients $u_1$ and $u_2$.
The eclipse model adds two additional free parameters per eclipse: the eclipse midpoint $T_{0,\mathrm{ecl}}$ and the depth $f_p$.
The minimization is first applied to the BI light curve for both targets, and then achromatic transit parameters are fixed for subsequent fits to the spectroscopic light curves as shown in Figures \ref{fig:light-curves-ltt} and \ref{fig:light-curves-wasp}.
While this approach to fitting planetary phase curves is relatively low-complexity and inappropriate for inferring physical parameters, it is appropriate for the purpose of removing the dominant planetary signatures to study the stellar noise.

\section{Model Selection}\label{section:appendix-model-selection}

We compare models which have different C (the Harvey Profile slope) and $N_{\rm components}$.
The standard Harvey profile parametrization using $C = 2$ corresponds in the time domain to a simple harmonic oscillator, while $C = 4$ corresponds to a symmetric rising and decaying pulse \citep{Michel2009,Kallinger2010A&A...522A...1K}.
Several studies have instead set $C = 4$ or have it as a free parameter \citep[e.g., ][]{Michel2009,Kallinger2010b,Kallinger2014A&A...570A..41K,Guo2022}.
In order to determine the best parameterization of N and $C$, we fit a suite of different models to either star's BI PSD. We vary N between one-component and two-components, and vary $C$ to be fixed at 2, fixed at 4, and in the case of two-components, all permutations of 2 and 4 ($\{2,2\}$, $\{4,4\}$, $\{2,4\}$, and $\{4,2\}$); a total of 8 models.
We then compare both the Bayesian Information Criteria (BIC) and Akaike Information Criteria (AIC) for the entire suite of models. 
We performed a series of tests in order to determine how many components and what values of $C$ are justified by these data.
Table \ref{tab:model-comparison} shows the results of this test, from which we conclude that a single component with C=2 (\modelone) is the best choice for measuring and reporting results for these systems.
Comparing all results across targets, the next best model appears to be N=2 with C=$\{2,4\}$ (\modelthree).

\subsection{Solar Comparison}
\begin{figure*}
    \centering
    \includegraphics[width=0.95\linewidth]{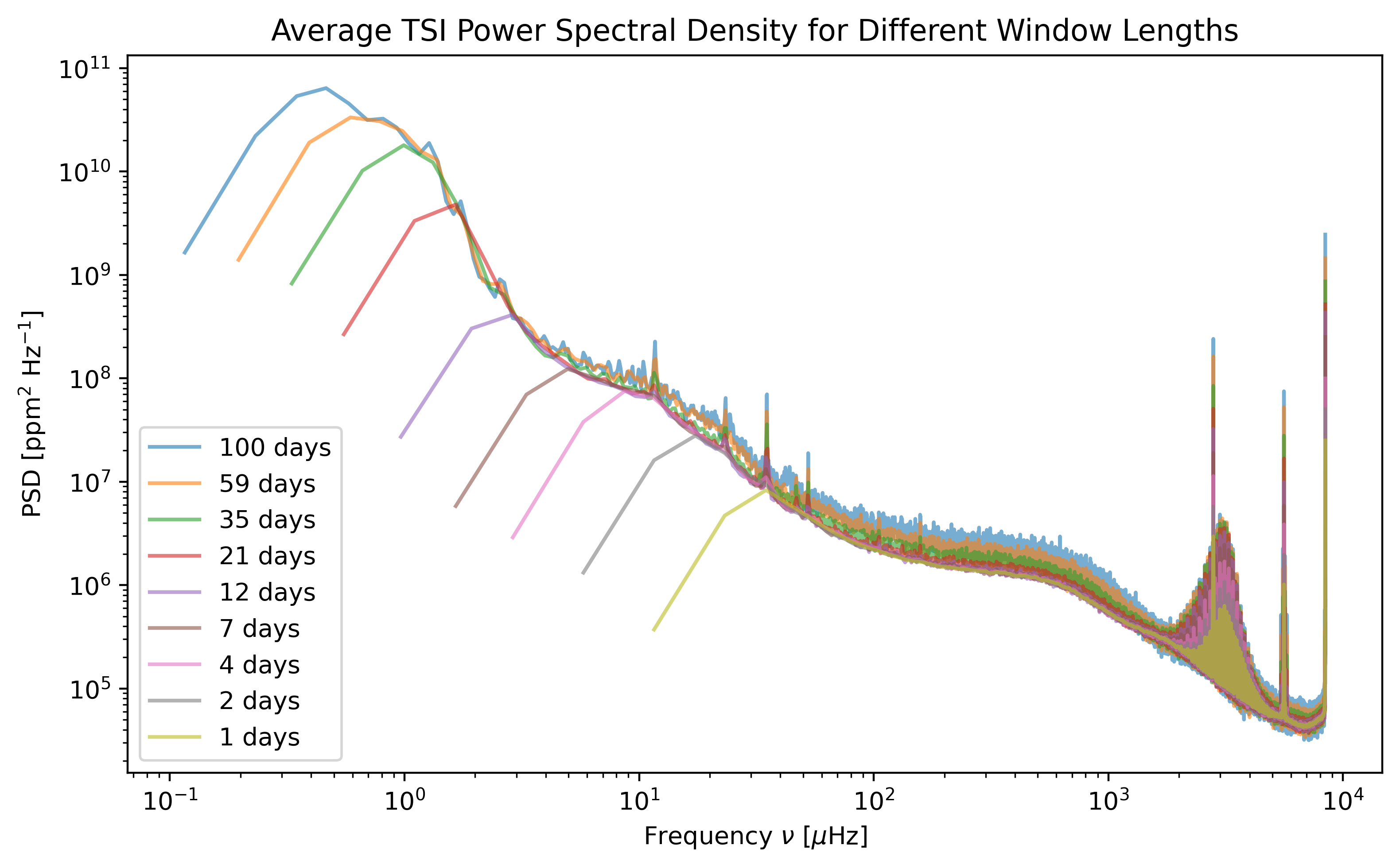}
    \caption{TSI PSD calculated using different windows to show how the low-frequency plateau is a feature of time series length. Each window is a number of days over which the PSD is calculated before being averaged across N chunks of the given window size, where N is the full length of the TSI photometry divided by window size.}
    \label{fig:appendixfig-solar-psd}
\end{figure*}

These model comparisons are not performed for the TSI data, as the Solar PSD is unambiguously best fit by a single component with C=2.
Figure \ref{fig:appendixfig-solar-psd} shows that, while a rise in power at lower frequencies is a genuine physical signature in the TSI data, the amplitude and frequency at which the ``plateau'' occurs is highly dependent on the length of the window from which the power spectrum is calculated.
This illustrates the sensitivity of the inferred low-frequency structure to the observing window and limits direct comparisons with the shorter JWST observations.
Based on this comparison, we do not expect to see any genuine evidence of a second granulation component on the timescale of these JWST observations.
We have computed and shown results of a ``second component'' for LTT-9779 and WASP-121 but this is not a granulation component -- rather, it is likely to be from some combination of stellar activity, residual planetary signatures, instrumental noise, or other sources of excess noise.
The solar comparison alone does not establish the physical origin of an additional component in the JWST spectra. We therefore treat the longer-timescale component of \modelthree~as a phenomenological description of excess low-frequency power, without identifying it as a separate granulation signal.

\begin{table*}
    \centering
    
    \begin{tabular}{l c c c c c c c}
        \toprule
        {Model} & k & {AIC} & {$\Delta$AIC} & $p(\mathrm{model})_{\rm AIC}$ & {BIC} & {$\Delta$BIC} & $p(\mathrm{model})_{\rm BIC}$ \\
        \midrule
        \multirow{1}{*}{LTT-9779} (n$_{\rm data}$=2016) & & & & & & & \\
        N=1, C=2 (\modelone)        &      6&     57330.08&       0.00&      0.5716 &    57363.74 &      0.00 &     0.9747 \\
        N=2, C=$\{2,4\}$ (\modelthree)       &      8&     57332.53&       2.45&      0.1679 &    57377.41 &     13.67 &     0.0010\\
        N=2, C=$\{2,2\}$          &      8&     57332.96&       2.88&      0.1353 &    57377.84 &     14.10 &     0.0008\\
        N=2, C=$\{4,2\}$         &      8&     57334.71&       4.63&      0.0565 &    57379.58 &     15.85 &     0.0004\\
        N=2, C=$\{4,4\}$          &      8&     57334.75&       4.67&      0.0555 &    57379.62 &     15.88 &     0.0003\\
        N=1, C=4 (\modeltwo)          &      6&     57337.60&       7.52&      0.0133 &    57371.26 &      7.52 &     0.0227 \\
        \midrule
        \multirow{1}{*}{WASP-121} (n$_{\rm data}$=1096) & & & & & & \\
        N=2, C=$\{2,4\}$ (\modelthree)       &      8&     32462.48&       0.00&      0.3410 &    32502.48 &      9.07 &     0.0105\\
        N=2, C=$\{2,2\}$          &      8&     32463.34&       0.86&      0.2219 &    32503.34 &      9.93 &     0.0068\\
        N=1, C=2 (\modelone)    &      6&     32463.41&       0.93&      0.2140 &    32493.41 &      0.00 &     0.9745\\
        N=2, C=$\{4,4\}$          &      8&     32464.51&       2.03&      0.1235 &    32504.51 &     11.10 &     0.0038\\
        N=2, C=$\{4,2\}$         &      8&     32464.95&       2.47&      0.0992 &    32504.95 &     11.54 &     0.0030\\
        N=1, C=4 (\modeltwo)      &      6&     32476.45&      13.97&      0.0003 &    32506.46 &     13.04 &     0.0014\\
        \bottomrule
    \end{tabular}
    \caption{Model comparison statistics. Reported values are based on the results of fitting the BI power spectrum and listed in descending model significance according to $\Delta$AIC. 
    }
    \label{tab:model-comparison}
\end{table*}

\section{Derivation of $\sigma_{\rm GP}$}\label{section:appendix-derivation-of-sigmagp}

The time series variance of the power spectrum used in \texttt{celerite} is defined as 
\begin{equation}
    \sigma^2_{\rm GP}=\frac{1}{\sqrt{2\pi}}\int_{-\infty}^{\infty}P(\omega)d\omega.
\end{equation}
Excluding negative frequencies and substituting eqn. \ref{eqn:sho_gp_pomega} for $P(\omega)$, we can rewrite this as:
\begin{equation}
    \sigma^2_{\rm GP}=\frac{2}{\pi}\int_{0}^{\infty}\frac{S_0}{1+(\omega/\omega_0)^4}d\omega.
\end{equation}
Solving this integral gives us:
\begin{equation}
    \sigma^2_{\rm GP}=\frac{S_0\omega_0}{\sqrt{2}}.
\end{equation}

Now we return to the definition of
$a_{\mathrm{gran}}$, the measured parameter reported as granulation amplitude in \citet{Coulombe2025} and \citet{Splinter2025}, which is 
\begin{equation}
    a_{\mathrm{gran,}GP}=\sqrt{\sqrt{2}S_0\omega_0}.
\end{equation}
Comparing these two equations, we find the following:
\begin{equation}
   \sigma^2_{\mathrm{gran},GP}=\frac{1}{2}a_{\mathrm{gran,}GP}^2. 
\end{equation}
This equation tells us how to relate the time series variance to the GP-reported amplitude, $a_{\mathrm{gran,}GP}$:
\begin{equation}
    \sigma_{\mathrm{gran},GP}=\frac{a_{\mathrm{gran,}GP}}{\sqrt{2}}.
\end{equation}

\end{document}